\documentclass[]{article}

\usepackage[utf8]{inputenc}
\usepackage{amsmath, amssymb, amsfonts}
\usepackage{amsthm}
\usepackage{bm}
\usepackage{graphicx}
\usepackage{fancyhdr}
\usepackage{titlesec}
\usepackage{indentfirst}
\usepackage{booktabs}
\usepackage{verbatim}
\usepackage{xcolor}
\usepackage{subcaption}

\usepackage[page,header]{appendix}
\usepackage{titletoc}
\usepackage{relsize}
\usepackage[colorlinks=true, linkcolor=blue, citecolor=blue, urlcolor=blue]{hyperref}

\allowdisplaybreaks 

\newcommand{\rd}{\,\mathrm{d}}

\newcommand{\bv}{\mathbf{v}}
\newcommand{\bx}{\mathbf{x}}

\newcommand{\bB}{\mathbf{B}}
\newcommand{\bE}{\mathbf{E}}
\newcommand{\bu}{\mathbf{u}}
\newcommand{\bU}{\mathbf{U}}
\newcommand{\bJ}{\mathbf{J}}

\newcommand{\bl}{\mathbf{l}}
\newcommand{\bA}{\mathbf{A}}
\newcommand{\bb}{\mathbf{b}}
\newcommand{\bc}{\mathbf{c}}
\newcommand{\bM}{\mathbf{M}}
\newcommand{\mQ}{\mathcal{Q}}
\newcommand{\blambda}{\bm{\lambda}}

\numberwithin{equation}{section}
\newtheorem{theorem}{Theorem}[section]

\newtheorem{remark}[theorem]{Remark}

\begin{document}
	
	\title{An Explicit Energy-Conserving Particle Method for the\\
		Vlasov--Fokker--Planck Equation}
	
	\author{Jiyoung Yoo\footnote{Department of Applied Mathematics, University of Washington, Seattle, WA 98195, USA (jyyoo496@uw.edu).}, \ 
		Jingwei Hu\footnote{Corresponding author. Department of Applied Mathematics, University of Washington, Seattle, WA 98195, USA (hujw@uw.edu).}, \
		Lee F. Ricketson\footnote{Lawrence Livermore National Laboratory, Center for Applied Scientific Computing, Livermore, CA 94550, USA (ricketson1@llnl.gov).}}
	
	\maketitle
	
\begin{abstract}
We propose an explicit particle method for the Vlasov--Fokker--Planck equation that conserves energy at the fully discrete level. The method features two key components: a deterministic and conservative particle discretization for the nonlinear Fokker--Planck operator (also known as the Lenard--Bernstein or Dougherty operator), and a second-order explicit time integrator that ensures energy conservation through an accuracy-justifiable correction. We validate the method on several plasma benchmarks, including collisional Landau damping and two-stream instability, demonstrating its effectiveness. 
\end{abstract}

\section{Introduction}

Plasmas are the most common form of matter, comprising more than 99\% of the visible universe. Understanding their complex behaviors has led to important advances in fields ranging from space
science and astronomy to the design of energy-generation devices such as fusion reactors, high-power microwave generators, and large particle accelerators.

Mathematically, plasmas are ionized gases whose evolution is well described by kinetic equations, a mesoscopic description of interacting-particle systems \cite{Villani02}. A widely used kinetic model for plasma dynamics is the Vlasov--Fokker--Planck equation (written here in dimensionless form):
\begin{equation}\label{eq:VFP}
    \partial_t f + \bv \cdot \nabla_{\bx}f + q(\bE + \bv \times \bB) \cdot \nabla_{\bv}f = \nu \mQ(f), \quad \bx\in \Omega_{\bx}\subset \mathbb{R}^d, \ \bv \in \mathbb{R}^d,
\end{equation}
where \( f = f(t, \bx, \bv) \) is the distribution function of some plasma species, depending on time $t$, position $\bx$, and velocity $\bv$, and $q = \pm 1$ is the normalized species charge. The right-hand side of (\ref{eq:VFP}) accounts for particle collisions via a nonlinear Fokker--Planck operator, with \( \nu \) representing the collision frequency and $\mQ(f)$ given by
\begin{equation} \label{eq:Qf}
\mQ(f)= \nabla_{\bv} \cdot \left(T\nabla_{\bv} f + (\bv - \bu)f \right),
\end{equation}
where \(T = T(t,\bx)\) is the temperature and \(\bu = \bu(t,\bx)\) is the bulk velocity. They are defined via the mass density $n(t,\bx) = \int_{\mathbb{R}^d} f(t,\bx, \bv)\rd{\bv}$ as
\begin{equation} \label{eq:uT}
\bu(t,\bx) = \frac{1}{n(t,\bx)}\int_{\mathbb{R}^d}  f(t,\bx, \bv) \bv \rd{\bv}, \quad T(t,\bx) = \frac{1}{d n(t,\bx)} \int_{\mathbb{R}^d}  f(t,\bx, \bv) |\bv - \bu(t,\bx)|^2 \rd{\bv}.
\end{equation}
In the physics literature, the Fokker--Planck operator is often referred to as the Lenard--Bernstein \cite{LB58} or Dougherty operator \cite{Doughterty64}, a simplified model introduced to mimic the full Landau collision operator \cite{Landau37}. Indeed, both operators share the following conservation property:
\begin{equation} \label{eq:cons-local}
\int_{\mathbb{R}^d} \mQ(f)\rd{\bv}=\int_{\mathbb{R}^d} \mQ(f)\bv\rd{\bv}=\int_{\mathbb{R}^d} \mQ(f)|\bv|^2\rd{\bv} = 0,
\end{equation}
as well as the Boltzmann's H-theorem:
\begin{equation*}
\int_{\mathbb{R}^d}\mQ(f) \log f\rd{\bv}\leq 0,
\end{equation*}
where the equality holds if and only if $f$ achieves the Maxwellian equilibrium:
\begin{equation*}
M[f]=\frac{n}{(2\pi T)^{d/2}}\exp\left(-\frac{|\bv-\bu|^2}{2T} \right).
\end{equation*}

Finally, \(\bE = \bE(t,\bx)\) and \(\bB= \bB(t,\bx)\) in equation \eqref{eq:VFP} are the electric and magnetic fields. In the fully electromagnetic case, they are determined self-consistently by the Maxwell’s equations:
\begin{align}
    \nabla_{\bx} \cdot \bE &= \rho - \rho_i, \label{Gauss's law}\\
    \nabla_{\bx} \cdot \bB &= 0, \label{Gauss's law for Mag}\\
   \partial_t\bE &=   \nabla_{\bx} \times \bB -\bJ, \label{Ampere-Maxwell law}\\
       \partial_t \bB &=- \nabla_{\bx} \times \bE , \label{Faraday's law}
\end{align}
where $\rho=\rho(t,\bx)$ and $\bJ=\bJ(t,\bx)$ are the charge density and current density defined by
\begin{equation} \label{eq:charge-current}
\rho(t,\bx) = q\int_{\mathbb{R}^d} f(t,\bx,\bv)\rd{\bv}, \quad \bJ(t,\bx) = q\int_{\mathbb{R}^d}   f(t,\bx, \bv)\bv\rd{\bv}.
\end{equation}
Here, $\rho_i$ in Gauss's law \eqref{Gauss's law} denotes a background density, assumed fixed on time-scales of interest, which serves to make the system globally neutral. It is given by $\rho_i = \frac{1}{|\Omega_{\bx}|} \int_{\Omega_{\bx}} \rho \rd{\bx},$
where $|\Omega_{\bx}|$ is the volume of the physical domain $\Omega_{\bx}$. 

In the electrostatic case, $\bB=\bB^{\text{ext}}(t,\bx)$ is given externally, and the electric field $\bE$ is curl free, hence can be represented as
\begin{equation*}
    \bE = - \nabla_{\bx} \phi,
\end{equation*}
where $\phi(t,\bx)$ is the electric potential. Substituting this form into Gauss’s law \eqref{Gauss's law} yields the Poisson's equation:
\begin{equation} \label{Poisson}
    -\Delta_{\bx} \phi = \rho - \rho_i.
\end{equation}
Alternatively, the electric field $\bE$ can be evolved with the Amp\`{e}re's law:
\begin{equation} \label{Ampere}
\partial_t \bE=-\bJ.
\end{equation}

Equation \eqref{eq:VFP} possesses many important physical properties. One particular property we are concerned with in this work is the energy conservation, which can be seen as follows: multiplying \eqref{eq:VFP} by $|\bv|^2/2$ and integrating in $\bv$ yields (using \eqref{eq:cons-local})
\begin{equation} \label{eq:cons-1}
\partial_t \int_{\mathbb{R}^d} \frac{1}{2}|\bv|^2 f \rd{\bv} +\nabla_{\bx} \cdot \int_{\mathbb{R}^d} \frac{1}{2}\bv |\bv|^2 f\rd{\bv}=\bJ\cdot \bE.
\end{equation}
From the Maxwell's equations, one can deduce
\begin{equation} \label{eq:cons-2}
\frac{1}{2}\partial_t \left(|\bE|^2+ |\bB|^2\right)=-\bJ\cdot \bE-\nabla_{\bx}\cdot (\bE\times \bB).
\end{equation}
Adding \eqref{eq:cons-1} and \eqref{eq:cons-2} and integrating in $\bx$ (assuming periodic boundary) gives 
\begin{equation} \label{eq:cons-global}
\partial_t \left( \int_{\Omega_{\bx}\times \mathbb{R}^d} \frac{1}{2}|\bv|^2 f\rd{\bv}\rd{\bx}+\frac{1}{2}\int_{\Omega_{\bx}} \left(|\bE|^2+|\bB|^2\right)\rd{\bx}\right)=0.
\end{equation}

Numerically solving equation \eqref{eq:VFP} is computationally expensive due to its potentially high dimensionality (total dimension is $2d$, where $d=1,2$, or 3). As such, the predominant numerical approach is the particle-in-cell (PIC) method \cite{BL, HE}. Based on tracing characteristics, this method naturally handles the Vlasov part of the equation (collisionless case), since it is a Hamiltonian flow. However, approximating the Fokker--Planck operator with particles is typically performed as a separate step using Monte Carlo-type methods, which introduces statistical noise \cite{bobylev2000theory, manheimer1997langevin, nanbu1997theory, TA77}. Moreover, these methods have first-order temporal accuracy at best, and in some cases their temporal accuracy is not well understood \cite{wang2008particle}.  Motivated by the recently introduced particle method for the Vlasov--Landau equation \cite{BCH24}, we propose in this work a deterministic particle method for the Fokker--Planck operator that can be naturally coupled with the classical PIC framework. We emphasize that, unlike the Landau operator, the conservation property \eqref{eq:cons-local} for the Fokker--Planck operator cannot be easily achieved in a particle setting. One contribution of our work is therefore a conservative particle approximation that guarantees this property. Having established the particle method, we further seek a time discretization that ensures energy conservation \eqref{eq:cons-global} of the full system. Since the collision term is the most computationally expensive part of the simulation, our goal is to design an explicit scheme. This is a highly non-trivial task; in fact, only recently was such a second-order scheme \cite{RH25} proposed by two of the authors for the collisionless case. The second contribution of our work is to extend this idea to the collisional system, including both electrostatic and electromagnetic cases. To summarize, we develop an explicit particle method for the full Vlasov--Fokker--Planck--Maxwell system that coherently incoporates transport, field effects, and collisions, while achieving energy conservation at the fully discrete level.

The rest of this paper is organized as follows. Section~\ref{Sec:particle} describes the particle method for the Vlasov--Fokker--Planck equation and, in particular, introduces a conservative particle approximation for the Fokker--Planck collision operator. Section~\ref{Sec:time-discretization} presents the energy-conserving time discretization for both the electrostatic and electromagnetic cases. Section~\ref{Sec:numerical} provides extensive numerical results that validate the proposed scheme. The paper is concluded in Section~\ref{Sec:conclusion}.

\section{Particle method for the Vlasov--Fokker--Planck equation}
\label{Sec:particle}

In the particle method \cite{Raviart85, Chertock}, the distribution function $f(t,\bx, \bv)$ is approximated by a collection of representative particles:
\begin{equation} 
\label{eq:fN}
	f^N(t,\bx, \bv) = \sum_{p=1}^{N} w_p \delta(\bx - \bx_p(t)) \delta(\bv - \bv_p(t)), 
\end{equation}
where $N$ is the number of particles, and \( w_p \), \( \bx_p(t) \), and \( \bv_p(t) \) denote the weight, position, and velocity of the \( p \)-th particle, respectively. 

To enable a particle method for the collision term \eqref{eq:Qf}, we first rewrite it as
\begin{equation*}
	\nabla_{\bv} \cdot (T\nabla_{\bv} f + (\bv - \bu)f) =\nabla_{\bv} \cdot \left(f \left(T\nabla_{\bv} \log f + (\bv - \bu)\right)\right)=\nabla_{\bv}\cdot (f \mathcal{U}[f]),
\end{equation*}
where 
\begin{equation} \label{eq:UU}
    \mathcal{U}[f](t,\bx,\bv):= T\nabla_{\bv} \log f + (\bv - \bu).
\end{equation}
This formulation interprets the Fokker--Planck operator as a continuity equation with a nonlinear velocity field $\mathcal{U}[f]$. Thus, it can be viewed as a forcing term similar to the field term in the Vlasov equation. This idea was already used in the 1990s \cite{DM90} to treat diffusion equations.

With this reformulation, the original equation \eqref{eq:VFP} can be written as
\begin{equation*} 
    \partial_t f +  \nabla_{\bx} \cdot (\bv f) + q  \nabla_{\bv}\cdot ((\bE + \bv \times \bB) f) = \nu \nabla_{\bv} \cdot \left( \mathcal{U}[f] f \right).
\end{equation*}
Hence, a particle method can be formally derived by tracing the characteristics:
\begin{align}
\begin{aligned}
    \frac{ \rd  \bx_p}{\rd t} &= \bv_p, \\
    \frac{\rd \bv_p}{ \rd t} &= q \left(\bE(t,\bx_p) + \bv_p \times \bB(t,\bx_p)\right) -\nu \mathcal{U}[f^N](t,\bx_p, \bv_p). \label{eq:v-particle}
    \end{aligned}
\end{align}
Note that since the equation is written in conservative form, the particle weights $w_p$ remain constant over time. The initial values \( \bx_p(0) \) and \( \bv_p(0) \) are obtained from sampling from the initial distribution function $f(0,\bx,\bv)$. 

\subsection{Regularization and interpolation with the kernel}

The particle solution \( f^N(t,\bx,\bv) \) in \eqref{eq:fN}, which is a sum of Dirac-delta functions, is not well defined in a form suitable for numerical computation. To properly define the electric and magnetic fields $\bE(t,\bx_p)$, $\bB(t,\bx_p)$, and collision term $\mathcal{U}[f^N](t,\bx_p, \bv_p)$ in \eqref{eq:v-particle}, we regularize $f^N$ as
\begin{equation} \label{eq:rfN}
 	f^N_{\eta,\epsilon} (t,\bx, \bv) = \sum_{p=1}^{N} w_p S_{\eta}(\bx - \bx_p(t)) S_{\epsilon}(\bv - \bv_p(t)),
 \end{equation}
where the kernel function $S_{\eta}(\bx)$ (and similarly $S_{\epsilon}(\bx)$) is defined as 
\begin{equation*}
S_{\eta}(\bx) = \prod_{i=1}^d \frac{1}{\eta^d}S\left(\frac{x_i}{\eta}\right),
\end{equation*}
and satisfies
\begin{equation*}
S_{\eta}(\bx)\geq 0, \quad \int_{\mathbb{R}^d}S_{\eta}(\bx)\rd{\bx}=1,\quad S_{\eta}(-\bx)=S_{\eta}(\bx).
\end{equation*}
Common base kernels $S(x)$ include the Gaussian $G(x)=\frac{1}{\sqrt{\pi}}e^{-x^2}$ and B-splines $B^m(x)$, where $m$ is the degree of the spline with $m=1$ corresponding to the tent function $B^1(x)=\max\{0,1-|x|\}$. 

Given the particle positions and velocities $\{\bx_p(t),\bv_p(t)\}_{p=1}^N$, if we wish to construct the moments of $f^N$---for example, the charge density and current density \eqref{eq:charge-current}---they can be approximated as
\begin{equation*}
    \rho(t,\bx) \approx q \sum_{p=1}^{N} w_p\, S_{\eta}(\bx - \bx_p(t)), \quad \bJ(t,\bx) \approx q \sum_{p=1}^{N} w_p\, \bv_p(t) S_{\eta}(\bx - \bx_p(t)).
\end{equation*}
On the other hand, if we are given quantities defined on regular grid points $\{\bx_h\}$ with mesh size $h$ in each dimension---such as $\{\bE_h(t)\}_h$ and $\{\bB_h(t)\}_h$ obtained from solving the Maxwell's equations---we can construct their interpolation using the $B^1$ spline (higher-degree B-splines may also be used but require solving a linear system to obtain the interpolation coefficients):
\begin{equation*}
    \bE(t,\bx) \approx \sum_{h} \bE_h(t)\, S_{\eta}(\bx - \bx_h)\eta^d, \quad \bB(t,\bx) \approx \sum_{h} \bB_h(t)\, S_{\eta}(\bx - \bx_h)\eta^d.
\end{equation*}
In this particular case, $\eta$ is identical to $h$.

Based on the above discussion, for the reminder of this paper we will always take $S_{\eta}$ in the physical space to be the $B^1$ spline for both regularization and interpolation (hence $\eta=h$). In the velocity space, we will always take $S_{\epsilon}$ to be the Gaussian kernel.

\subsection{Conservative particle approximation for the Fokker--Planck operator}

Given the regularization \eqref{eq:rfN} and the definition \eqref{eq:UU}, it is natural to approximate the velocity field $\mathcal{U}[f^N](t,\bx_p,\bv_p)$ as follows:
\begin{equation} \label{eq:velocity-field}
\mathcal{U}[f^N](t,\bx_p,\bv_p) \approx T_p\nabla_{\bv} \log f^N_{h,\epsilon}(\bx_p,\bv_p) + (\bv_p - \bu_p):=\bU(\bx_p,\bv_p),
\end{equation}
where 
\begin{equation*}
    \nabla_{\bv} \log f^N_{h, \epsilon}(\bx_p,\bv_p) 
    =\nabla_{\bv} \log \left(\sum_{q=1}^{N} w_q S_h(\bx_p - \bx_q) S_{\epsilon}(\bv_p - \bv_q)\right) = \frac{\sum_{q=1}^N w_q S_h(\bx_p - \bx_q) \nabla_{\bv}S_{\epsilon}(\bv_p-\bv_q)}{\sum_{q=1}^N w_q S_h(\bx_p - \bx_q)S_{\epsilon}(\bv_p-\bv_q)}. 
\end{equation*}
The next critical step is to determine $T_p$ and $\bu_p$. From their definition in \eqref{eq:uT}, a first approximation can be 
\begin{equation} \label{eq:temp_first}
   \overline{\bu}_p = \frac{1}{\overline{n}_p}\sum_{q=1}^{N} w_q \bv_q S_{h}(\bx_p - \bx_q), \quad \overline{T}_p = \frac{1}{d\overline{n}_p}\sum_{q=1}^{N} w_q |\bv_q - \overline{\bu}_p|^2 S_{h}(\bx_p - \bx_q),
    \end{equation}
where  $\overline{n}_p := \sum_{q=1}^{N} w_q S_{h}(\bx_p - \bx_q)$.
However, such defined $\overline{T}_p$ and $\overline{\bu}_p$ do not conserve momentum and energy as in the continuous model (cf.~\eqref{eq:cons-local}). To achieve this, we propose to determine $T_p$ and $\bu_p$ by solving the optimization problem:
\begin{align}
    &\min_{ \{T_p, \bu_p\}_{p=1}^N} \ \sum_{p=1}^N \left( |T_p - \overline{T}_p|^2 + |\bu_p- \overline{\bu}_p|^2 \right) \label{eq:opt problem}\\
    & \quad \text{s.t. } \sum_{p=1}^N w_p \bU(\bx_p,\bv_p)=0, \quad \sum_{p=1}^N w_p \bv_p \cdot \bU(\bx_p,\bv_p)=0,  \label{eq:constraints}
\end{align}
where the constraints \eqref{eq:constraints} correspond to, respectively, momentum and energy conservation.

Define $\bl_p := \nabla_{\bv} \log f_{h, \epsilon}^N(\bx_p, \bv_p)$, then \eqref{eq:constraints} can be written as
\begin{equation}
\label{eq:constaints1}
\sum_{p=1}^N w_p \left(T_p\bl_p + \bv_p - \bu_p\right)=0, \quad \sum_{p=1}^N w_p \bv_p \cdot \left(T_p\bl_p + \bv_p - \bu_p\right)=0.
\end{equation}
Introducing the Lagrange multipliers $\blambda_1\in \mathbb{R}^d$, $\lambda_2\in \mathbb{R}$, we obtain
\begin{align*}
    \mathcal{L}(T_1,\dots,T_N,\bu_1,\dots,\bu_N, \blambda_1, \lambda_2) &= \sum_{p=1}^N \left(|T_p - \overline{T}_p|^2 + |\bu_p - \overline{\bu}_p|^2\right) \\ 
    &+ \blambda_1 \cdot \left( \sum_{p=1}^N w_p \left(T_p\bl_p + \bv_p - \bu_p\right)\right) + \lambda_2 \left(\sum_{p=1}^N w_p \bv_p \cdot \left(T_p\bl_p + \bv_p - \bu_p\right)\right).
\end{align*}
Computing $\partial_{T_p} \mathcal{L}$ gives
\begin{equation*}
2(T_p-\overline{T}_p)+ w_p\blambda_1\cdot \bl_p+ \lambda_2 w_p \bv_p \cdot \bl_p=0.
\end{equation*}
Computing $\partial_{\bu_p} \mathcal{L}$ gives
\begin{equation*}
2\left(\bu_p-\overline{\bu}_p\right)-\blambda_1 w_p-\lambda_2 w_p \bv_p=0.
\end{equation*}
Then we have
\begin{equation} \label{eq:modified T}
T_p=\overline{T}_p-\frac{1}{2}w_p\blambda_1\cdot \bl_p-\frac{1}{2}\lambda_2 w_p \bv_p\cdot \bl_p,
\end{equation}
and
\begin{equation} \label{eq:modified u}
\bu_p=\overline{\bu}_p+ \frac{1}{2}w_p\blambda_1+\frac{1}{2}\lambda_2 w_p \bv_p.
\end{equation}
Substituting $T_p$ and $\bu_p$ into \eqref{eq:constaints1} gives
\begin{align*}
    &\sum_{p=1}^N w_p\left[\left(\overline{T}_p-\frac{1}{2}w_p\blambda_1\cdot \bl_p-\frac{1}{2}\lambda_2 w_p \bv_p\cdot \bl_p\right) \bl_p + \bv_p - \left(\overline{\bu}_p+ \frac{1}{2}w_p\blambda_1+\frac{1}{2}\lambda_2 w_p \bv_p\right)\right] =0,\\
    &\sum_{p=1}^N w_p \bv_p \cdot \left[\left(\overline{T}_p-\frac{1}{2}w_p\blambda_1\cdot \bl_p-\frac{1}{2}\lambda_2 w_p \bv_p\cdot \bl_p\right) \bl_p + \bv_p - \left(\overline{\bu}_p+ \frac{1}{2}w_p\blambda_1+\frac{1}{2}\lambda_2 w_p \bv_p\right) \right] = 0,
\end{align*}
which can be rewritten as a linear system in terms of $\blambda_1$ and $\lambda_2$ as follows:
\begin{align} \label{eq:linear-system}
\begin{aligned} 
 &\left[\frac{1}{2}\sum_{p=1}^N w_p^2 \left(\bl_p \otimes \bl_p + \mathbf{I} \right) \right] \blambda_1 +  \left[\frac{1}{2}\sum_{p=1}^N w_p^2 \left(\bl_p \otimes \bl_p + \mathbf{I} \right) \bv_p \right] \lambda_2 = \sum_{p=1}^N w_p \left(\overline{T}_p \bl_p + \bv_p - \overline{\bu}_p\right), \\
 & \left[\frac{1}{2}\sum_{p=1}^N w_p^2 \left(\bl_p \otimes \bl_p + \mathbf{I} \right) \bv_p\right] \cdot \blambda_1 + \left[\frac{1}{2}\sum_{p=1}^N w_p^2 \bv_p^T \left(\bl_p \otimes \bl_p + \mathbf{I}\right) \bv_p \right] \lambda_2 = \sum_{p=1}^N w_p \bv_p \cdot \left(\overline{T}_p \bl_p + \bv_p - \overline{\bu}_p\right),
 \end{aligned}
\end{align}
where $\bl_p \otimes \bl_p$ denotes the tensor product and $\mathbf{I}$ is the $d \times d$ identity matrix. We next show that this linear system indeed has a unique solution. Once $\blambda_1$ and $\lambda_2$ are obtained, they can be substituted into equations \eqref{eq:modified T} and \eqref{eq:modified u} to recover the solution to the optimization problem. 

Introducing the notation
\begin{align*}
  &\bM_p =\frac{1}{2}w_p^2\left(\bl_p \otimes \bl_p + \mathbf{I}\right), \quad   \bA = \sum_{p=1}^N \bM_p, \quad \bb = \sum_{p=1}^N \bM_p \bv_p, \quad e = \sum_{p=1}^N \bv_p^T\bM_p \bv_p,\\
    & \bc = \sum_{p=1}^N w_p \left(\overline{T}_p \bl_p + \bv_p - \overline{\bu}_p\right), \quad g = \sum_{p=1}^N w_p \bv_p \cdot \left(\overline{T}_p \bl_p + \bv_p - \overline{\bu}_p\right),
\end{align*}
then the linear system \eqref{eq:linear-system} can be formulated as
\begin{align*}
    \begin{bmatrix}
        \bA & \bb \\
        \bb^T & e
    \end{bmatrix} \begin{bmatrix}
        \blambda_1 \\ \lambda_2
    \end{bmatrix} = 
    \begin{bmatrix}
        \bc \\ g
    \end{bmatrix},
\end{align*}
which has a solution if and only if the following two conditions hold:
\begin{equation*}
   \bA \succ 0,  \quad s := e - \bb^T \bA^{-1} \bb \neq 0,
\end{equation*}
where $s$ is the Schur complement of $\bA$. First, since $\bM_p$ is positive definite for each $p$, and $\bA$ is a sum of such positive definite matrices, $\bA$ is also positive definite and hence invertible. 
Now define a vector $\hat{\bv}:= \bA^{-1} \bb$, then 
\begin{align*}
    s& = e - \bb^T \bA^{-1} \bb= e - \hat{\bv}^T\bA\hat{\bv} = \sum_{p=1}^N \bv_p^T\bM_p \bv_p - \sum_{p=1}^N \hat{\bv}^T\bM_p \hat{\bv}  \\
    &= \sum_{p=1}^N \left(\bv_p - \hat{\bv}\right)^T \bM_p \left(\bv_p - \hat{\bv}\right) -2\hat{\bv}^T \underbrace{\left(\sum_{p=1}^N\bM_p \right)}_{= \ \bA}\hat{\bv} + 2 \hat{\bv}^T \underbrace{\left(\sum_{p=1}^N \bM_p \bv_p\right)}_{= \ \bb \ = \ \bA \hat{\bv} } \\
    &= \sum_{p=1}^N \left(\bv_p - \hat{\bv}\right)^T \bM_p \left(\bv_p - \hat{\bv}\right)\geq 0,
\end{align*}
where the equality holds if and only if $\bv_p = \hat{\bv}$, $\forall p= 1, \dots, N$,  which implies all $\bv_p$ are identical. Since this does not occur in practice, we conclude that $s>0$.

\begin{remark}
One thing we cannot justify is the positivity of $T_p$ obtained from the above optimization procedure. Note that $\overline{T}_p$, as defined in \eqref{eq:temp_first}, is positive. Therefore, if the correction in \eqref{eq:modified T} is not large, the resulting $T_p$ should remain positive. In all the numerical examples presented in this paper, we did not observe any negative $T_p$ values.
\end{remark}

\begin{remark}
In the spatially homogeneous case, $T$ and $\bu$ are independent of particles, so the above procedure is significantly simplified. We seek $T$ and $\bu$ such that the following constraints hold: 
\begin{align} \label{eq:homo-system}
\begin{aligned}
  \sum_{p=1}^N w_p \bU(\bv_p) &= \sum_{p=1}^N w_p \left(T \bl_p + \bv_p - \bu\right) = 0, \\
    \sum_{p=1}^N w_p \bv_p \cdot \bU(\bv_p) &= \sum_{p=1}^N w_p \bv_p \cdot \left(T \bl_p + \bv_p - \bu\right) = 0,
    \end{aligned}
\end{align}
where
\begin{equation*}
\bl_p:=\nabla_\bv \log f_{\epsilon}^N(\bv_p)=
    \nabla_{\bv} \log \left(\sum_{q=1}^{N} w_q  S_{\epsilon}(\bv_p - \bv_q)\right) = \frac{\sum_{q=1}^N w_q \nabla_{\bv}S_{\epsilon}(\bv_p-\bv_q)}{\sum_{q=1}^N w_q S_{\epsilon}(\bv_p-\bv_q)}. 
\end{equation*}
The linear system \eqref{eq:homo-system} can be rearranged as follows:
\begin{align*}
    &\left(\sum_{p=1}^N w_p \bl_p\right)T - \left(\sum_{p=1}^N w_p\right) \bu = - \sum_{p=1}^N
 w_p \bv_p, \\
 &\left(\sum_{p=1}^N w_p \bv_p \cdot \bl_p\right)T - \left(\sum_{p=1}^N w_p \bv_p\right) \cdot \bu = - \sum_{p=1}^N
 w_p |\bv_p|^2. 
\end{align*}
In order for it to have a solution, it suffices to check that
\begin{equation*}
 \left(\sum_{p=1}^N w_p\right)\left(\sum_{p=1}^N w_p \bv_p \cdot \bl_p\right) -\left(\sum_{p=1}^N w_p \bl_p\right) \cdot \left(\sum_{p=1}^N w_p \bv_p\right)\neq 0.
\end{equation*}
If we assume $f^N_{\epsilon}(\bv)=\sum_{q=1}^{N} w_q  S_{\epsilon}(\bv - \bv_q)$ is $\lambda$-strongly log-concave, i.e., $\nabla^2 \log f^N_{\epsilon} \preceq -\lambda I$, then the left hand side of the above inequality can be rearranged to yield
\begin{align*}
\frac{1}{2}\sum_{p,q=1}^N w_p w_q (\bv_p-\bv_q)\cdot (\bl_p-\bl_q)=\frac{1}{2}\sum_{p,q=1}^N w_p w_q (\bv_p-\bv_q)\cdot \nabla^2 \log f^N_{\epsilon}(\xi_{pq})(\bv_p-\bv_q)\leq -\frac{\lambda}{2}\sum_{p,q=1}^N w_p w_q |\bv_p-\bv_q|^2,
\end{align*}
which is only zero if all $\bv_p$ are identical. In general, we do not have a rigorous argument for this, but the linear system always admits a solution in all the homogeneous tests in this paper.

We note that this special case is similar to the approach introduced recently in \cite{JKHP24}. However, our optimization procedure for determining $T_p$ and $\bu_p$ in the spatially inhomogeneous case is more general.
\end{remark}

\begin{remark}
An alternative way of approximating the velocity field is
\begin{equation*} 
\mathcal{U}[f^N](t,\bx_p,\bv_p) \approx T_p\nabla_{\bv}\frac{\delta H_{h,\epsilon}}{\delta f}(\bx_p,\bv_p) + (\bv_p - \bu_p),
\end{equation*}
where 
\begin{equation*}
\nabla_{\bv}\frac{\delta H_{h,\epsilon}}{\delta f}(\bx_p,\bv_p) =\frac{\sum_{q=1}^N w_q S_{h}(\bx_p - \bx_q) \nabla_{\bv}S_{\epsilon}(\bv_p-\bv_q)}{\sum_{q=1}^N w_q S_{h}(\bx_p - \bx_q)S_{\epsilon}(\bv_p-\bv_q)}+\sum_{q=1}^N w_q \frac{S_{h}(\bx_p - \bx_q) \nabla_{\bv}S_{\epsilon}(\bv_p-\bv_q)}{\sum_{r=1}^N w_r S_{h}(\bx_q - \bx_r)S_{\epsilon}(\bv_q-\bv_r)}.
\end{equation*}
This formulation guarantees that, in the spatially homogeneous case, the entropy defined by
\begin{equation*}
H_{\epsilon}:=\sum_{p=1}^N w_p \log\left(\sum_{q=1}^N S_{\epsilon}(\bv_p-\bv_q)\right)
\end{equation*}
satisfies the decay property:
\begin{align*}
\frac{\rd H_{\epsilon}}{\rd t}&=\sum_{p=1}^N w_p \nabla_{\bv}\frac{\delta H_{\epsilon}}{\delta f}(\bv_p)\cdot \frac{\rd \bv_p}{\rd t}=\sum_{p=1}^N w_p \left(\nabla_{\bv}\frac{\delta H_{\epsilon}}{\delta f}(\bv_p)+\frac{\bv_p-\bu}{T}\right)\cdot \frac{\rd \bv_p}{\rd t}\\
&=-\sum_{p=1}^N w_p \left(\nabla_{\bv}\frac{\delta H_{\epsilon}}{\delta f}(\bv_p)+\frac{\bv_p-\bu}{T}\right)\cdot T \left(\nabla_{\bv}\frac{\delta H_{\epsilon}}{\delta f}(\bv_p)+\frac{\bv_p-\bu}{T}\right)\leq 0.
\end{align*}
The second equality in the first line above can be achieved by imposing $T$, $\bu$ such that
\begin{equation*}
\sum_{p=1}^N w_p \left( T \nabla_{\bv}\frac{\delta H_{\epsilon}}{\delta f}(\bv_p)+ (\bv_p-\bu)\right) = 0, \quad \sum_{p=1}^N w_p \bv_p\cdot \left( T \nabla_{\bv}\frac{\delta H_{\epsilon}}{\delta f}(\bv_p)+ (\bv_p-\bu)\right) = 0,
\end{equation*}
which are the same type of constraints as in \eqref{eq:homo-system}.

This line of reasoning has led to the development of entropy-decaying particle methods in a series of papers (e.g., \cite{CCP19, CHWW20, BCH24}). Since our main focus in this work is on conservation properties, we adopt the simpler choice given in \eqref{eq:velocity-field}.
\end{remark}

\section{Energy-conserving time discretization}
\label{Sec:time-discretization}

We have now obtained a semi-discrete particle method for the Vlasov--Fokker--Planck equation \eqref{eq:VFP}:
\begin{align} \label{eq:particle-system}
\begin{aligned}
    \frac{ \rd  \bx_p}{\rd t} &= \bv_p, \\
    \frac{\rd \bv_p}{ \rd t} &= q \left(\bE(t,\bx_p) + \bv_p \times \bB(t,\bx_p)\right) -\nu \bU(\bx_p,\bv_p),
 \end{aligned}
\end{align}
where $\bU(\bx_p,\bv_p)$ is given by \eqref{eq:velocity-field}. Although $T_p$ and $\bu_p$ in the definition of $\bU(\bx_p,\bv_p)$ are determined from the optimization problem \eqref{eq:opt problem} to ensure conservation, there is no guarantee that, once time discretization is applied to the particle system \eqref{eq:particle-system}, the resulting scheme will preserve the energy as in the continuous model \eqref{eq:cons-global}. In fact, as reported in \cite{BCH24}, the energy may grow proportionally to the time step $\Delta t$ if a naive forward Euler scheme is used. Such a violation of energy conservation can be problematic, leading to significant deviations from the true dynamics or even instability in long-time simulations. 

Therefore, we aim to design a time discretization that preserves the energy at the fully discrete level. Moreover, since evaluating the collision term  $\bU(\bx_p,\bv_p)$ constitutes the most expensive part of the simulation, we seek a time integrator that is fully explicit. Since the collision effect is often weak in many plasma applications, an explicit scheme is not expected to suffer from severe stability constraints.

Designing an explicit energy-conserving PIC scheme (even without collisions) is a highly non-trivial task. Only recently did two of the authors proposed such a scheme \cite{RH25}. The basic idea is to first construct an explicit midpoint PIC scheme and then apply an accuracy-justifiable correction to each individual particle to enforce energy conservation. In what follows, we show that the same idea can be generalized to the collisional system, while keeping the collision term fully explicit. We begin with the electrostatic case and present two versions of the scheme. Both are second-order and energy-conserving, but version 2 improves upon version 1 by providing a more robust correction. We will then discuss the generalization to the electromagnetic case. 

Before proceeding, we note that, for the proposed time discretization to work, the only property we require of the Fokker--Planck part is the conservation property \eqref{eq:constraints}. This property holds readily for the particle method for the Landau collision operator introduced in \cite{BCH24}. Therefore, the time integration schemes presented in this work can be used in conjunction with the particle method of \cite{BCH24} to achieve energy conservation.

\subsection{Electrostatic case: version 1 scheme}

Assume that, at the beginning of time step $t^n$, $\bx_p^n$, $\bv_p^n$, and $\bE_h^n$ (the electric field defined on the regular grid points $\bx_h$) are available. We propose the following scheme:
\begin{equation} \label{version1}
\begin{aligned}
    \bx_p^* &= \bx_p^n + \frac{\Delta t}{2} \bv_p^n, \\
    \bv_p^* &= \bv_p^n + \frac{\Delta t}{2} q\left( \bE_p^{n,*} + \bv_p^* \times \bB^{\text{ext}}(t^{n+\frac{1}{2}},\bx_p^*) \right)-\frac{\Delta t}{2}\nu \bU(\bx_p^*,\bv_p^n), \\
    \bx_p^{n+1} &= \bx_p^n + \Delta t\, \bv_p^*, \\  \bE^{n+1}_h &= \bE^n_h - \Delta t\, \bJ_h^{*,*}, \\
    \bv_p^{\dagger} &= \bv_p^n + \Delta t q \left( \bE_p^{n+\frac{1}{2},*} + \bv_p^* \times \bB^{\text{ext}}(t^{n+\frac{1}{2}},\bx_p^*) \right)-{\Delta t}\nu \bU(\bx_p^*,\bv_p^*), \\
    \bv_p^{n+1} &= \Gamma_p^n \bv_p^{\dagger},
\end{aligned}
\end{equation}
where 
\begin{align*}
    \bE_p^{n,*} &= \sum_{h} \bE_h^n S_h(\bx_p^* - \bx_h)h^d, \quad &\bJ_h^{*,*} &= q \sum_{p=1}^{N} w_p \bv_p^* S_h(\bx_h-\bx_p^*), \\
    \bE_p^{n+\frac{1}{2},*} &= \sum_{h} \bE_h^{n+\frac{1}{2}} S_h(\bx_p^* - \bx_h)h^d, \quad &\bE_h^{n+\frac{1}{2}} &= \frac{\bE_h^n + \bE_h^{n+1}}{2},
\end{align*}
and
\begin{align}\label{Gamma_p^n}
\Gamma_p^n = \sqrt{1+2 \frac{\left(\bv_p^{\dagger} - \bv_p^n \right) \cdot \left(\bv_p^* - \frac{\bv_p^{\dagger}+\bv_p^n}{2}\right) }{\left |\bv_p^{\dagger}\right|^2}}.
\end{align}

Note that, up to the step that computes $\bv_p^{\dagger}$, the scheme \eqref{version1} is essentially a second-order explicit midpoint method. However, energy is not conserved because
\begin{align} \label{energy-conservation}
   \sum_{p=1}^{N} w_p \mathbf{v}_p^* \cdot \left(\mathbf{v}_p^{\dagger} - \mathbf{v}_p^n \right)
    &= \Delta t \sum_{p=1}^N w_p \bv_p^* \cdot \left(q \left(\bE_p^{n+\frac{1}{2},*} + \bv_p^* \times \bB^{\text{ext}}\left(t^{n+\frac{1}{2}}, \bx_p^*\right)\right) - \nu \bU\left(\bx_p^*, \bv_p^*\right)\right) \notag\\
    &= \Delta t \sum_{p=1}^N w_p \bv_p^* \cdot q \bE_p^{n+\frac{1}{2},*} \quad \left(\because \sum_{p=1}^N w_p \bv_p^* \cdot \bU(\bx_p^*, \bv_p^*) = 0\right)  \notag\\ 
    &= \Delta t \sum_{p=1}^N w_p \bv_p^* \cdot q \left(\sum_h\bE_h^{n+\frac{1}{2}} S_h\left(\bx_p^* - \bx_h\right) h^d\right) \notag\\
    &= \Delta t \sum_h \bE_h^{n+\frac{1}{2}} h^d \cdot \left(q\sum_{p=1}^N w_p \bv_p^* S_h\left(\bx_h - \bx_p^*\right)\right) \notag\\
    &= \Delta t h^d \sum_h \bE_h^{n+\frac{1}{2}}  \cdot \bJ_h^{*,*}  \notag  \\
    &= \Delta t h^d \sum_h \bE_h^{n+\frac{1}{2}}  \cdot \left(-\frac{\bE_h^{n+1} - \bE_h^n}{\Delta t}\right) \notag\\
    &= - \frac{h^d}{2} \sum_h \left(\bE_h^{n+1} +  \bE_h^{n}\right) \cdot\left(\bE_h^{n+1} - \bE_h^{n}\right) \notag\\
    &= - \frac{h^d}{2} \sum_h \left(\left| \bE_h^{n+1} \right|^2 - \left |\bE_h^n\right|^2\right).
\end{align}
Therefore, the last step in \eqref{version1} serves to correct each individual particle to enforce energy conservation. In fact, it can be easily verified that $\bv_p^{n+1}$ satisfies
\begin{equation} \label{eq:cons}
\frac{1}{2} \left(|\bv_p^{n+1}|^2 - |\bv_p^n|^2 \right)=\bv_p^* \cdot \left(\bv_p^{\dagger} - \bv_p^n \right).
\end{equation}
Combining \eqref{eq:cons} with \eqref{energy-conservation} yields
\begin{equation*}
    \frac{1}{2} \sum_{p=1}^{N} w_p |\mathbf{v}_p^{n+1}|^2 + \frac{h^d}{2} \sum_h \left| \bE_h^{n+1} \right|^2  = \frac{1}{2} \sum_{p=1}^{N} w_p |\mathbf{v}_p^{n}|^2 + \frac{h^d}{2} \sum_h \left| \bE_h^{n} \right|^2,
\end{equation*}
i.e., the total energy is conserved at each time step.

It remains to show that the correction step does not destroy the second-order accuracy of the overall scheme. It suffices to show that 
\begin{equation} \label{version1:correction}
\Gamma_p^n = \sqrt{1+O(\Delta t^3)}=1+O(\Delta t^3).
\end{equation}

We assume all terms in \eqref{Gamma_p^n} are smooth and $O(1)$. Clearly,
\[
\bv_p^{\dagger}-\bv_p^n=O(\Delta t),
\]
and
\begin{align*}
    \bv_p^* - \frac{\bv_p^{\dagger}+\bv_p^n}{2} =& \ \bv_p^n +\frac{\Delta t}{2} q\left( \bE_p^{n,*} + \bv_p^* \times \bB^{\text{ext}}(t^{n+\frac{1}{2}},\bx_p^*) \right)-\frac{\Delta t}{2}\nu \bU(\bx_p^*,\bv_p^n) \\
    & \ -\frac{1}{2}\left(2\bv_p^n  + \Delta t q \left( \bE_p^{n+\frac{1}{2},*} + \bv_p^* \times \bB^{\text{ext}}(t^{n+\frac{1}{2}},\bx_p^*) \right)-{\Delta t}\nu \bU(\bx_p^*,\bv_p^*)\right) \\
    =& \ \frac{\Delta t}{2} q\underbrace{\left(\bE_p^{n,*} - \bE_p^{n+\frac{1}{2},*}\right)}_{(a)} -\frac{\Delta t}{2}\nu \underbrace{\left(\bU\left(\bx_p^*, \bv_p^n\right) - \bU\left(\bx_p^*, \bv_p^*\right)\right)}_{(b)}.
\end{align*}
For term $(a)$, we have
\begin{align*}
    \bE_p^{n,*} - \bE_p^{n+\frac{1}{2},*} &= \sum_h \bE_h^n S_h\left(\bx_p^* - \bx_h\right)h^d - \sum_h \frac{\bE_h^{n+1} + \bE_h^n}{2} S_h\left(\bx_p^* - \bx_h\right) h^d \\
    &= \frac{1}{2} \sum_h \left(\bE_h^n - \bE_h^{n+1}\right) S_h\left(\bx_p^* - \bx_h\right)h^d \\
    &= \frac{\Delta t}{2} \sum_h \bJ_h^{*,*} S_h\left(\bx_p^* - \bx_h\right) h^d \\
    &= O(\Delta t).
\end{align*}
For term $(b)$, we apply the Taylor expansion in the $\bv$ direction:
\begin{align*}
        \bU\left(\bx_p^*, \bv_p^*\right) &=  \bU\left(\bx_p^*, \bv_p^n + \frac{\Delta t}{2} q\left( \bE_p^{n,*} + \bv_p^* \times \bB^{\text{ext}}(t^{n+\frac{1}{2}},\bx_p^*) \right)-\frac{\Delta t}{2}\nu \bU(\bx_p^*,\bv_p^n)\right) \\
        &= \bU\left(\bx_p^*, \bv_p^n\right) + \left[\nabla_{\bv}\bU(\bx, \bv)\right]_{(\bx_p^*, \bv_p^n)} \left(\frac{\Delta t}{2} q\left( \bE_p^{n,*} + \bv_p^* \times \bB^{\text{ext}}(t^{n+\frac{1}{2}},\bx_p^*) \right)-\frac{\Delta t}{2}\nu \bU(\bx_p^*,\bv_p^n)\right) \\
        &= \bU\left(\bx_p^*, \bv_p^n\right) + O(\Delta t),
\end{align*}
where $\nabla_{\bv}\bU(\bx, \bv)  \in \mathbb{R}^{d\times d} $ is the Jacobian matrix of $\bU\left(\bx, \bv\right)$. We then have
\begin{align*}
    \bU\left(\bx_p^*, \bv_p^n\right)- \bU\left(\bx_p^*, \bv_p^*\right) = O(\Delta t).
\end{align*} 
Therefore, 
\begin{align*}
    \bv_p^* - \frac{\bv_p^{\dagger}+\bv_p^n}{2} = O(\Delta t^2).
\end{align*}
Altogether, we have shown \eqref{version1:correction}.

The quantity $\Gamma_p^n$ is computed at each time step for every particle, but it is not necessarily real. When $\Gamma_p^n$ is imaginary, we set it to $1$ so 
that the overall accuracy of the scheme is unaffected. Such problematic particles are rare; however, when they do occur, they may affect energy conservation. To further reduce their occurrence, an improved version is introduced in the next subsection. 

\subsection{Electrostatic case: version 2 scheme}

Assume that, at the beginning of time step $t^n$, $\bx_p^n$, $\bv_p^n$, and $\bE_h^n$ are available. We propose the following improved scheme:
\begin{equation}  \label{version2}
\begin{aligned}
\bx_p^* &= \bx_p^n + \frac{\Delta t}{2} \bv_p^n, \\
\bv_p^{**} &= \bv_p^n + \frac{\Delta t}{2} q \left(\bE_p^{n,*} + \bv_p^{**} \times \bB^{\text{ext}}\left(t^{n+\frac{1}{2}}, \bx_p^*\right)\right) - \frac{\Delta t}{2}\nu \bU\left(\bx_p^*, \bv_p^n\right), \\
\bE^*_h &= \bE^n_h - \frac{\Delta t}{2} \bJ^{**,*}_h, \\
\bv_p^* &= \bv_p^n + \frac{\Delta t}{2}q \left(\bE_p^{*,*} + \bv_p^* \times \bB^{\text{ext}}\left(t^{n+\frac{1}{2}}, \bx_p^*\right)\right) - \frac{\Delta t}{2}\nu \bU\left(\bx_p^*, \bv_p^{**}\right), \\
\bx_p^{n+1} &= \bx_p^n + \Delta t \bv_p^*, \\
\bE^{n+1}_h &= \bE^n_h - \Delta t \bJ^{*,*}_h, \\
\bv_p^{\dagger} &= \bv_p^n + \Delta t q  \left(\bE_p^{n+\frac{1}{2},*} + \bv_p^* \times \bB^{\text{ext}}\left(t^{n+\frac{1}{2}}, \bx_p^*\right)\right) - \Delta t\nu \bU\left(\bx_p^*, \bv_p^*\right), \\
\bv_p^{n+1} &= \Gamma_p^n \bv_p^{\dagger},
\end{aligned}
\end{equation}
where 
\begin{align*}\bE_p^{*,*} = \sum_{h} \bE_h^* S_h(\bx_p^* - \bx_h)h^d, \quad 
    \bJ^{**,*}_h = q\sum_{p=1}^N w_p \bv_p^{**} S_h\left(\bx_h - \bx_p^*\right),
\end{align*}
and the other quantities (not repeated here) follow the same definitions as in the version 1 scheme.

Compared with the version 1 scheme, only lines 2-4 in \eqref{version2} are new. In particular, the argument that leads to energy conservation in the version 1 scheme applies verbatim here. However, we will show that these additional steps boost \eqref{Gamma_p^n} to (compare with \eqref{version1:correction})
\begin{equation} \label{version2:correction}
\Gamma_p^n = \sqrt{1+O(\Delta t^4)}=1+O(\Delta t^4).
\end{equation}
Indeed, we still have $\bv_p^{\dagger}-\bv_p^n=O(\Delta t)$, but we now claim that 
\begin{equation} \label{eq:boost}
\bv_p^* - \frac{\bv_p^{\dagger}+\bv_p^n}{2}=O(\Delta t^3)
\end{equation}
as shown below.
\begin{align*}
    \bv_p^* - \frac{\bv_p^{\dagger} + \bv_p^n}{2} =& \ \bv_p^n + \frac{\Delta t}{2} q\left(\bE_p^{*,*} + \bv_p^* \times \bB^{\text{ext}}\left(t^{n+\frac{1}{2}},\bx_p^*\right)\right) - \frac{\Delta t}{2}\nu \bU\left(\bx_p^*, \bv_p^{**}\right) \\
    & \ -\frac{1}{2} \left(2\bv_p^n + \Delta t q\left(\bE_p^{n+\frac{1}{2},*} + \bv_p^* \times \bB^{\text{ext}}\left(t^{n+\frac{1}{2}}, \bx_p^*\right)\right) - \Delta t \nu \bU\left(\bx_p^*, \bv_p^*\right)\right) \\
    =& \ \frac{\Delta t}{2} q \underbrace{\left(\bE_p^{*,*} - \bE_p^{n+\frac{1}{2},*} \right)}_{(c)} - \frac{\Delta t}{2} \nu \underbrace{\left(\bU\left(\bx_p^* , \bv_p^{**}\right)- \bU\left(\bx_p^*, \bv_p^*\right)\right)}_{(d)}.
\end{align*}
For term $(c)$, we have
\begin{align*}
     \bE_p^{*,*} - \bE_p^{n+\frac{1}{2},*} &= \sum_h \bE_h^* S_h \left(\bx_p^* - \bx_h\right)h^d - \sum_h \frac{\bE_h^{n+1} + \bE_h^n}{2}S_h\left(\bx_p^* - \bx_h\right) h^d \\
    &= \sum_h \left(\bE_h^n - \frac{\Delta t}{2}\bJ_h^{**,*} - \frac{\bE_h^{n+1} + \bE_h^n}{2}\right)S_h\left(\bx_p^* - \bx_h\right) h^d \\
    &= \frac{1}{2}\sum_h \left(\bE_h^{n} - \bE_h^{n+1} - \Delta t \bJ_h^{**,*}\right) S_h\left(\bx_p^* - \bx_h\right)h^d \\
    &= \frac{1}{2}\sum_h \left(\Delta t \bJ_h^{*,*} - \Delta t \bJ_h^{**,*}\right)S_h\left(\bx_p^* - \bx_h\right) h^d \\
    &= \frac{\Delta t}{2}\sum_h \left(q\sum_{p=1}^N w_p\underbrace{\left(\bv_p^* - \bv_p^{**} \right)}_{= \ O(\Delta t)}S_h\left(\bx_h - \bx_p^*\right) \right)S_h\left(\bx_p^* - \bx_h\right)h^d \\
    &= O(\Delta t^2).
\end{align*}
For term $(d)$, we first show that $\bv_p^{**} - \bv_p^* = O(\Delta t^2)$:
\begin{align*}
    \bv_p^{**} - \bv_p^* =& \ \frac{\Delta t}{2}q \underbrace{\left(\bE_p^{n,*} - \bE_p^{*,*}\right)}_{(d_1)} + \frac{\Delta t}{2}q \left(\underbrace{\left(\bv_p^{**} - \bv_p^*\right)}_{(d_2)}\times \bB^{\text{ext}}\left(t^{n+\frac{1}{2}}, \bx_p^*\right)\right)  - \frac{\Delta t}{2}\nu\underbrace{\left(\bU\left(\bx_p^*, \bv_p^n\right) - \bU\left(\bx_p^*, \bv_p^{**}\right)\right)}_{(d_3)}.
\end{align*}
First, for $(d_1)$, we have
\begin{align*}
    \bE_p^{n,*} - \bE_p^{*,*} &= \sum_h \left(\bE_h^* - \bE_h^n\right) S_h(\bx_p^* - \bx_h)\, h^d = -\frac{\Delta t}{2} \sum_h \bJ_h^{**,*} S_h(\bx_p^* - \bx_h)\, h^d = O(\Delta t).
\end{align*}
Term $(d_2)$ is $\bv_p^{**} - \bv_p^* = O(\Delta t)$, and term $(d_3)$ is $\bU(\bx_p^*, \bv_p^n) - \bU(\bx_p^*, \bv_p^{**}) = O(\Delta t)$ because $\bv_p^n - \bv_p^{**} = O(\Delta t)$.
Therefore, we can extract an additional factor of $\Delta t$ from $\bv_p^{**} - \bv_p^*$ and obtain $ \bv_p^{**} - \bv_p^* = O(\Delta t^2)$. This implies
\[
\bU(\bx_p^*, \bv_p^{**}) - \bU(\bx_p^*, \bv_p^*) = O(\Delta t^2),
\]
by applying the Taylor expansion in the $\bv$ direction.

Altogether, we have shown \eqref{eq:boost}, hence \eqref{version2:correction}.

\subsection{Extension to the electromagnetic case}

The schemes above for the electrostatic case can be easily extended to the electromagnetic case. In particular, a generalization of the version 2 scheme is given as follows.

Assume that, at the beginning of time step $t^n$, $\bx_p^n$, $\bv_p^n$, and $\bE_h^n$, $\bB_h^n$ are available. Then
\begin{equation} 
\begin{aligned}
\bx_p^* &= \bx_p^n + \frac{\Delta t}{2} \bv_p^n, \\
\bv_p^{**} &= \bv_p^n + \frac{\Delta t}{2} q \left(\bE_p^{n,*} + \bv_p^{**} \times \bB_p^{n,*}\right)  - \frac{\Delta t}{2}\nu \bU\left(\bx_p^*, \bv_p^n\right), \\
\bE^*_h &= \bE^n_h +\frac{\Delta t}{2}\left(\nabla_h\times \bB_h^n-\bJ^{**,*}_h\right), \\
\bB^*_h &= \bB^n_h -\frac{\Delta t}{2} \nabla_h\times \bE_h^n, \\
\bv_p^* &= \bv_p^n + \frac{\Delta t}{2}q \left(\bE_p^{*,*} + \bv_p^* \times \bB_p^{*,*}\right) - \frac{\Delta t}{2}\nu \bU\left(\bx_p^*, \bv_p^{**}\right), \\
\bx_p^{n+1} &= \bx_p^n + \Delta t \bv_p^*, \\
\bE^{n+1}_h &= \bE^n_h +\Delta t \left(\nabla_h\times \bB_h^{n+\frac{1}{2}} -\bJ^{*,*}_h\right), \\
\bB^{n+1}_h &= \bB^n_h - \Delta t \nabla_h\times \bE_h^{n+\frac{1}{2}}, \\
\bv_p^{\dagger} &= \bv_p^n + \Delta t q  \left(\bE_p^{n+\frac{1}{2},*} + \bv_p^* \times \bB_p^{n+\frac{1}{2},*}\right) - \Delta t\nu \bU\left(\bx_p^*, \bv_p^*\right), \\
\bv_p^{n+1} &= \Gamma_p^n \bv_p^{\dagger}.
\end{aligned}
\end{equation}
Note that $\bB_p^{n,*}$, $\bB_p^{*,*}$, $\bB_h^{n+1/2}$, and $\bB_p^{n+1/2,*}$ are defined in precisely the same way as the corresponding $\bE$-field quantities. This scheme is implicit in the field solve while maintaining the explicit particle push. 

To show the energy conservation, we note the following
\begin{align*}
    \sum_{p=1}^{N} w_p \mathbf{v}_p^* \cdot \left(\mathbf{v}_p^{\dagger} - \mathbf{v}_p^n \right)& = \Delta t \sum_{p=1}^N w_p \bv_p^* \cdot \left(q \left(\bE_p^{n+\frac{1}{2},*} + \bv_p^* \times \bB_p^{n+\frac{1}{2},*}\right) - \nu \bU\left(\bx_p^*, \bv_p^*\right)\right)\\
   &= \Delta t \sum_{p=1}^N w_p \bv_p^* \cdot q \bE_p^{n+\frac{1}{2},*}  
    = \Delta t \sum_{p=1}^N w_p \bv_p^* \cdot q \left(\sum_h\bE_h^{n+\frac{1}{2}} S_h\left(\bx_p^* - \bx_h\right) h^d\right) \\
    &= \Delta t \sum_h \bE_h^{n+\frac{1}{2}} h^d \cdot \left(q\sum_{p=1}^N w_p \bv_p^* S_h\left(\bx_h - \bx_p^*\right)\right) \\
    &= \Delta t h^d \sum_h \bE_h^{n+\frac{1}{2}}  \cdot \bJ_h^{*,*} = \Delta t h^d \sum_h \bE_h^{n+\frac{1}{2}}  \cdot \left(\nabla_h\times \bB_h^{n+\frac{1}{2}}-\frac{\bE_h^{n+1} - \bE_h^n}{\Delta t}\right) \\
    &= \Delta t h^d\sum_h \left( \left(\nabla_h\times \bE_h^{n+\frac{1}{2}}\right)\cdot \bB_h^{n+\frac{1}{2}}-\nabla_h\cdot \left( \bE_h^{n+\frac{1}{2}}\times \bB_h^{n+\frac{1}{2}}\right)\right) \\
    &\quad  - \frac{h^d}{2} \sum_h \left(\bE_h^{n+1} +  \bE_h^{n}\right) \cdot\left(\bE_h^{n+1} - \bE_h^{n}\right) \\
    &=   - \frac{h^d}{2} \sum_h \left(\bB_h^{n+1} +  \bB_h^{n}\right) \cdot\left(\bB_h^{n+1} - \bB_h^{n}\right)     - \frac{h^d}{2} \sum_h \left(\bE_h^{n+1} +  \bE_h^{n}\right) \cdot\left(\bE_h^{n+1} - \bE_h^{n}\right)         \\
    &= - \frac{h^d}{2} \sum_h \left(\left| \bB_h^{n+1} \right|^2 - \left |\bB_h^n\right|^2\right)- \frac{h^d}{2} \sum_h \left(\left| \bE_h^{n+1} \right|^2 - \left|\bE_h^n\right|^2\right).
\end{align*}
Combining this with \eqref{eq:cons} (which still holds since the correction step remains the same as in the electrostatic case) leads to
\begin{equation*}
    \frac{1}{2} \sum_{p=1}^{N} w_p |\mathbf{v}_p^{n+1}|^2 + \frac{h^d}{2} \sum_h \left| \bE_h^{n+1} \right|^2 +\frac{h^d}{2} \sum_h \left| \bB_h^{n+1} \right|^2 = \frac{1}{2} \sum_{p=1}^{N} w_p |\mathbf{v}_p^{n}|^2 + \frac{h^d}{2} \sum_h \left| \bE_h^{n} \right|^2+\frac{h^d}{2} \sum_h \left| \bB_h^{n} \right|^2.
\end{equation*}

By a similar procedure, we show that $\Gamma_p^n = \sqrt{1+O(\Delta t^4)}=1+O(\Delta t^4)$. We claim that 
\[
\bv_p^* - \frac{\bv_p^{\dagger}+\bv_p^n}{2}=O(\Delta t^3)
\]
as shown below.
\begin{align*}
    \bv_p^* - \frac{\bv_p^{\dagger} + \bv_p^n}{2} =& \frac{\Delta t}{2} q\left(\bE_p^{*,*} + \bv_p^* \times \bB_p^{*,*}\right) - \frac{\Delta t}{2}\nu \bU\left(\bx_p^*, \bv_p^{**}\right) -\frac{\Delta t}{2} q\left(\bE_p^{n+\frac{1}{2},*} + \bv_p^* \times \bB_p^{n+\frac{1}{2},*}\right) + \frac{\Delta t}{2} \nu \bU\left(\bx_p^*, \bv_p^*\right) \\
    =& \ \frac{\Delta t}{2} q \underbrace{\left(\bE_p^{*,*} - \bE_p^{n+\frac{1}{2},*} +\bv_p^*\times \left(\bB_p^{*,*}-\bB_p^{n+\frac{1}{2},*}\right)\right)}_{(c)} - \frac{\Delta t}{2} \nu \underbrace{\left(\bU\left(\bx_p^* , \bv_p^{**}\right)- \bU\left(\bx_p^*, \bv_p^*\right)\right)}_{(d)}.
\end{align*}
For term $(c)$, we have
\begin{align*}
  & \bE_p^{*,*} - \bE_p^{n+\frac{1}{2},*} +\bv_p^*\times \left(\bB_p^{*,*}-\bB_p^{n+\frac{1}{2},*}\right) \\
   =& \sum_h \left(\bE_h^*  -\bE_h^{n+\frac{1}{2}}\right)S_h\left(\bx_p^* - \bx_h\right)h^d +\sum_h \bv_p^*\times\left(\bB_h^*  -\bB_h^{n+\frac{1}{2}}\right)S_h\left(\bx_p^* - \bx_h\right)h^d \\
=& \frac{\Delta t}{2}\sum_h \left(\nabla_h\times \bB_h^n - \bJ_h^{**,*} -\nabla_h\times \bB_h^{n+\frac{1}{2}}+\bJ_h^{*,*} \right)S_h\left(\bx_p^* - \bx_h\right) h^d \\
& \quad +\frac{\Delta t}{2}\sum_h \bv_p^*\times \left(\nabla_h\times \bE_h^{n+\frac{1}{2}}-\nabla_h\times \bE_h^n\right)S_h\left(\bx_p^* - \bx_h\right)h^d\\
=& \frac{\Delta t}{2}\sum_h \left(\nabla_h\times \frac{\bB_h^n -\bB_h^{n+1}}{2}+\bJ_h^{*,*} - \bJ_h^{**,*} \right)S_h\left(\bx_p^* - \bx_h\right) h^d \\
&\quad +\frac{\Delta t}{2}\sum_h \bv_p^*\times \left(\nabla_h\times \frac{\bE_h^{n+1}-\bE_h^n}{2} \right)S_h\left(\bx_p^* - \bx_h\right) h^d\\
=& O(\Delta t^2).
\end{align*}
For term $(d)$, it suffices to show that $\bv_p^{**} - \bv_p^* = O(\Delta t^2)$. To see this, consider
\begin{align*}
    \bv_p^{**} - \bv_p^* =& \ \frac{\Delta t}{2}q \underbrace{\left(\bE_p^{n,*} - \bE_p^{*,*}\right)}_{(d_1)} + \frac{\Delta t}{2}q \left(\underbrace{\bv_p^{**}\times \bB_p^{n,*} - \bv_p^*\times \bB_p^{*,*}}_{(d_2)}\right) - \frac{\Delta t}{2}\nu\underbrace{\left(\bU\left(\bx_p^*, \bv_p^n\right) - \bU\left(\bx_p^*, \bv_p^{**}\right)\right)}_{(d_3)}.
\end{align*}
First, for $(d_1)$, we have
\begin{align*}
    \bE_p^{n,*} - \bE_p^{*,*} &= \sum_h \left(\bE_h^* - \bE_h^n\right) S_h(\bx_p^* - \bx_h)\, h^d= -\frac{\Delta t}{2} \sum_h \bJ_h^{**,*} S_h(\bx_p^* - \bx_h)\, h^d = O(\Delta t).
\end{align*}
Same can be argued for $\bB_p^{n,*}-\bB_p^{*,*}=O(\Delta t)$, hence term $(d_2)$ is $O(\Delta t)$ since $\bv_p^{**} - \bv_p^* = O(\Delta t)$, and term $(d_3)$ is $\bU(\bx_p^*, \bv_p^n) - \bU(\bx_p^*, \bv_p^{**}) = O(\Delta t)$.
Therefore, we can extract an additional factor of $\Delta t$ from $\bv_p^{**} - \bv_p^*$ and obtain $\bv_p^{**} - \bv_p^* = O(\Delta t^2)$.

\begin{remark}
We note that the linearly implicit temporal discretization of Maxwell's equations used here is but one of several possible choices.  In \cite{RH25}, it is shown that in the collisionless case two popular, fully explicit spatiotemporal discretizations of Maxwell's equations also yield exact energy conservation when used appropriately.  These are (a) the Yee lattice with a leapfrog temporal discreziation -- often called the FDTD method \cite{Yee66} -- and (b) the pseudospectral analytic time domain (PSATD) method \cite{Birdsall, vay2013domain}.  This analysis carries over trivially to the collisional case studied here, since the collisional term in the total energy vanishes by construction, so exact conservation can still be expected with these other Maxwell solvers.  
\end{remark}

\section{Numerical results}
\label{Sec:numerical}

In this section, we present extensive numerical results for the proposed energy-conserving particle method. Both version 1 and version 2 schemes are considered. We begin with the spatially homogeneous case to demonstrate the accuracy and energy-conserving property of the method for the collision part only. We then move to the spatially inhomogeneous case, presenting results for collisional linear and nonlinear Landau damping, as well as the collisional two-stream instability. Before that, we first discuss some implementation details used to accelerate the particle method.

\subsection{Cell-list optimization and GPU implementation}

In the evaluation of the collision term $\bU(\bx_p,\bv_p)$, both $\nabla_{\bv} \log f^{N}_{h,\epsilon}$ and $\overline{T}_p$ and $\overline{\bu}_p$ require particle-to-particle computations. These scale as $O(N^2)$ (where $N$ is the total number of particles, usually on the order of $10^5\sim10^6$) and constitute the dominant computational cost of the simulation. To address this, we employed two acceleration techniques: one at the algorithmic level and the other at the implementation level.

First, the cell-list optimization \cite{AT87} exploits the fact that interactions between distant particles are negligible because the kernels vanish beyond their support. Only nearby particles need to be considered, while distant pairs can be ignored without loss of accuracy. This can be done readily for the spline kernel $S_h$ in physical space, which has compact support. Although the Gaussian kernel $S_\epsilon$ in velocity space does not have compact support, we set the cell size in the $v$-domain to $6 \epsilon$ so that the resulting error remains around machine precision. This technique greatly reduces the number of kernel evaluations and achieves speed-ups by several tens compared with the naive full pairwise method. 

However, even with the cell-list optimization, the collision term $\bU(\bx_p,\bv_p)$ must still be evaluated for every particle at each time step. To further accelerate this, we compute the collision term on the GPU so that these evaluations can be performed in parallel. Simpler tasks, such as the field solve and particle push, still remain on the CPU, so data must be transferred between the CPU and GPU at each step. Despite this overhead, GPU parallelization significantly reduces the total simulation time.

\subsection{Spatially homogeneous case} 

In this subsection, we present the numerical results for the spatially homogeneous case, that is, no $\bx_p$-dependence in \eqref{eq:particle-system}, and the particle method reduces to
\begin{align} \label{eqn:homo-particle}
    \frac{\rd \bv_p}{ \rd t} = -\nu \bU(\bv_p).
\end{align}
The velocity space is assumed to be one-dimensional (1V).

\subsubsection{Order of convergence}\label{subsubsec:conv order}

We first validate the convergence order of the proposed particle method. Although no theoretical result current exists, we anticipate the method to exhibit an order of
\begin{equation} \label{eq:error}
O\left(\Delta v^2 + \frac{1}{\sqrt{N_c}}\right),
\end{equation}
where $\Delta v$ is the velocity mesh size and $N_c$ is the number of particles per velocity cell. We emphasize that the method itself does not use any mesh in the velocity space; $\Delta v$ is introduced only to define a suitable regularization parameter, $\epsilon=\Delta v$. The half-th order convergence arises from the use of Monte Carlo sampling for initialization. This order could potentially be improved with quasi-random or deterministic sampling, but we do not explore those options in the current work.

The initial distribution is chosen as a bimodal Gaussian:
\begin{align} \label{eqn:homo-bimodal}
	f_0(v) = \frac{1}{2\sqrt{2\pi}} \left( e^{-\tfrac{(v - 2.4)^2}{2}} + e^{-\tfrac{(v + 2.4)^2}{2}} \right),
\end{align}
which converges to a Maxwellian over time due to collisional effects. We set the collision strength to $\nu=0.05$, the time step to $\Delta t = 0.01$, and run the simulation up to the final time $t_{\text{final}} = 10$. The Gaussian kernel with $\epsilon = \Delta v$ is used in both the simulation and the reconstruction of the solution $f$. We choose the velocity grid numbers as $N_v = [8, \ 16, \ 32, \ 64, \ 128]$. The length of the velocity domain is defined as $L_v = \max_p v_p^0 - \min_p v_p^0$, where $v^0$ denotes the initial particle velocities, and the mesh size is then given by $\Delta v = L_v / N_v$. To ensure that the error in \eqref{eq:error} is dominated by the discretization in velocity space rather than by Monte Carlo sampling, we set $N_c$ sufficiently large. We first approximate $L_v \approx 12$, and with the finest grid $N_v = 128$, this gives the smallest mesh size $\Delta v = 12/128 = 0.09375$. The condition $\tfrac{1}{\sqrt{N_c}} < \Delta v^2$ then requires $N_c \geq 12946$. We set $N_c = 10000$, which does not strictly satisfy this bound but is sufficient to observe the expected order of convergence. 

\begin{figure}[htp!]
    \centering   \includegraphics[width=0.5\linewidth]{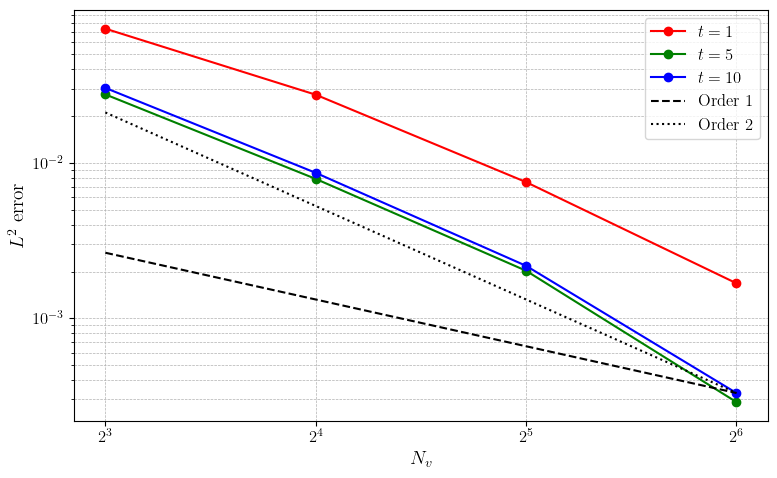}
    \caption{Spatially homogeneous test. Order of convergence: $L^2$ error relative to the finest mesh $N_v=128$.}
    \label{fig:conv_order}
\end{figure}

Since the version 1 and version 2 schemes exhibit the same convergence order, we present results only for the version 1 scheme. 
Figure~\ref{fig:conv_order} shows the $L^2$ errors relative to the finest mesh $N_v = 128$ at $t = 1,\,5,\,10$. The results confirm near second-order convergence in velocity space.

\subsubsection{Energy conservation property}\label{subsubsec:BM results}

We next demonstrate the energy conservation property of the version 1 and version 2 schemes. For reference, we compare them with the forward Euler time-stepping applied to the particle system \eqref{eqn:homo-particle}, which is known not to conserve energy. The initial distribution is the same as in \eqref{eqn:homo-bimodal}. We use $N_v = 64$ and $N_c = 16$, giving a total of $N = 1024$ particles. The simulations are run with a time step $\Delta t = 0.01$ up to $t_{\text{final}} = 10$. The Gaussian kernel with $\epsilon = \Delta v$ is used for both simulation and reconstruction, and the collision frequencies are set to $\nu = 0.01, \ 0.05, \ 0.1,$ and $0.15$.

\begin{figure}[htp!]
    \centering
    \begin{subfigure}[b]{0.48\textwidth}      \includegraphics[width=\linewidth]{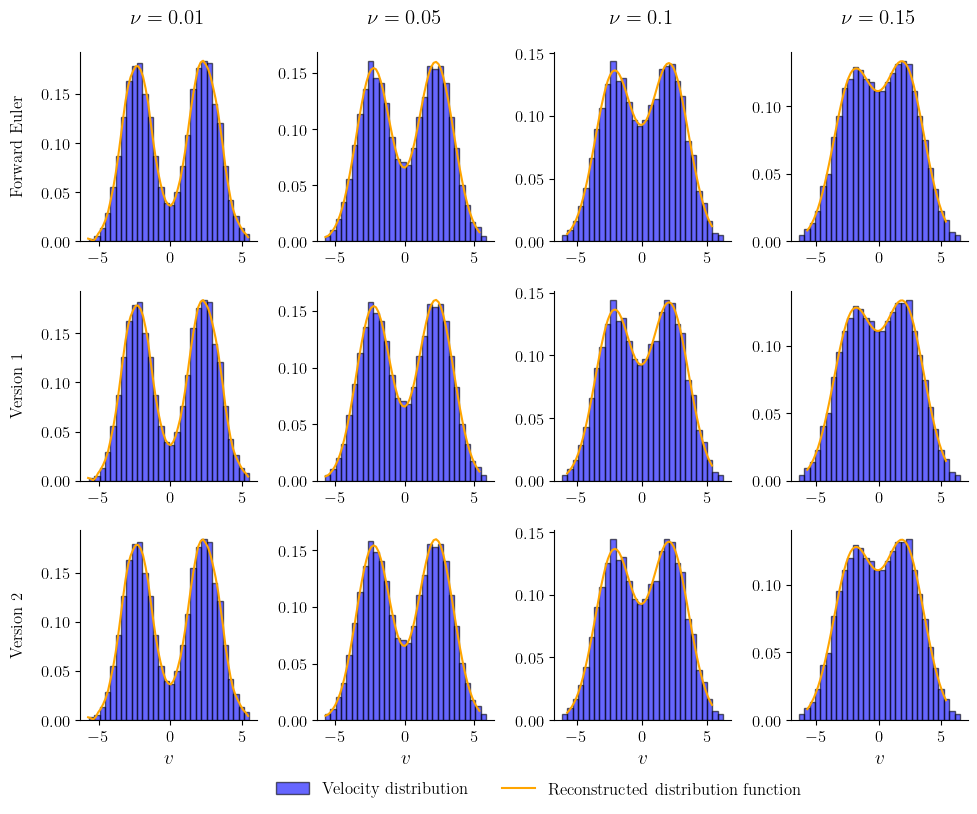}
        \caption{$t = 1$}
    \end{subfigure}
    \hfill
    \begin{subfigure}[b]{0.48\textwidth}
\includegraphics[width=\linewidth]{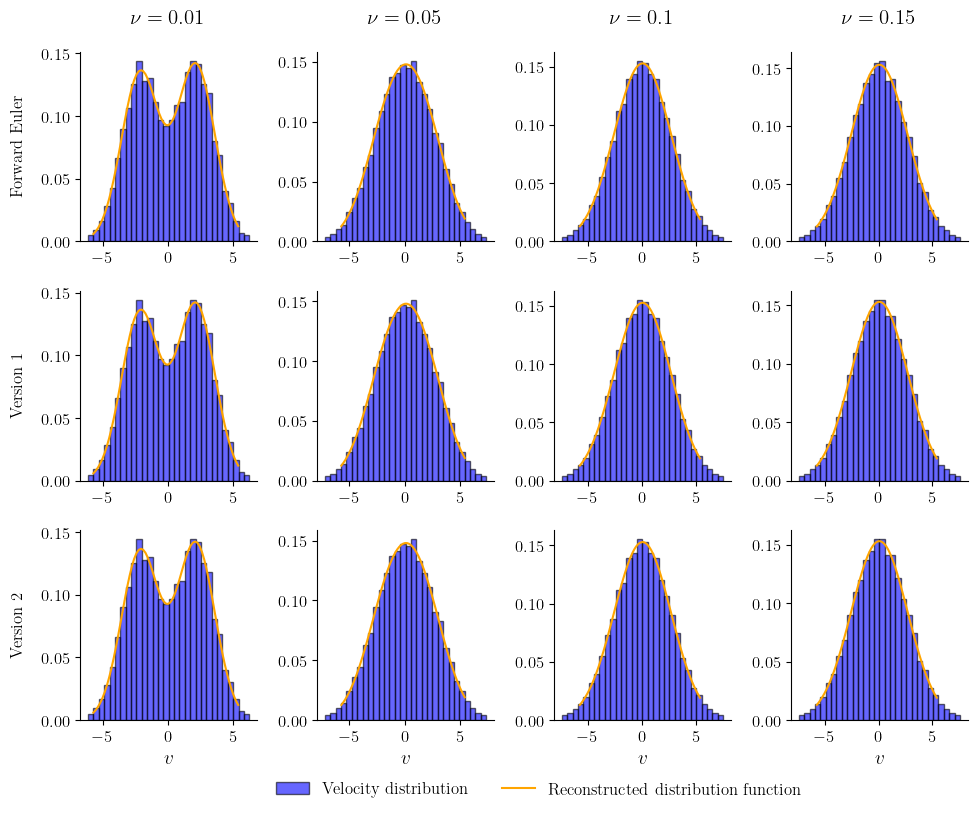}
        \caption{$t = 10$}
    \end{subfigure}
    \caption{Spatially homogeneous test. Velocity distributions at $t = 1$ and $10$ computed using the forward Euler, version 1, and version 2 schemes. Blue: histogram of the particle velocities. Orange: reconstructed distribution function. }
   \label{fig:BM_recons}
\end{figure}

Figure~\ref{fig:BM_recons} shows the time evolution of the velocity distribution with the forward Euler, version 1, and version 2 schemes. The bimodal distribution is driven toward the Maxwellian by collisions.  Stronger collisions make this trend faster. At this level, no significant difference is observed among all three schemes. 

\begin{figure}[htp!]
    \centering    \includegraphics[width=\linewidth]{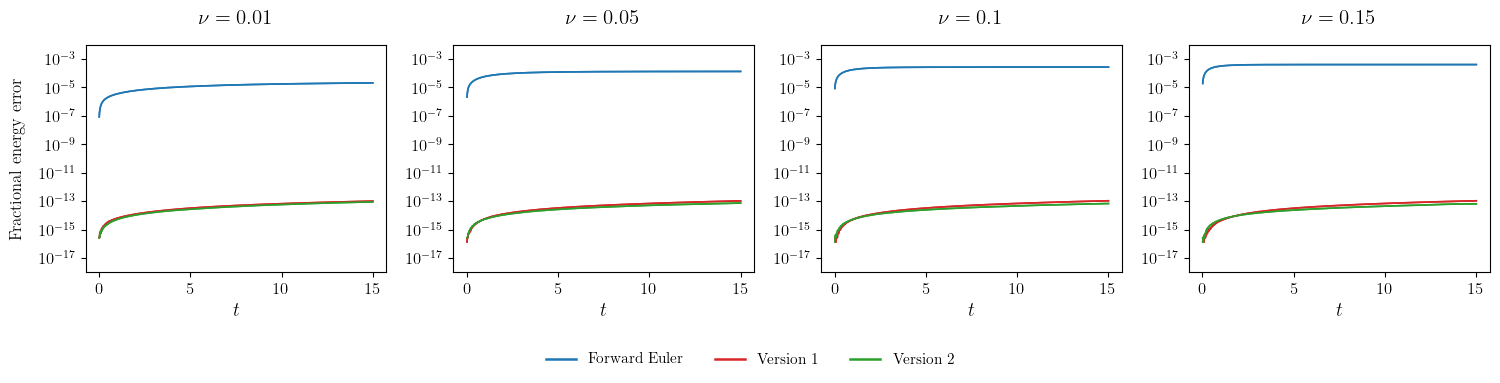}
    \caption{Spatially homogeneous test. Fractional change in total energy over time.}
    \label{fig:BM_fracE}
\end{figure}

Figure~\ref{fig:BM_fracE} presents the fractional change in total energy. Here, the total energy and its fractional change are defined as 
\begin{equation*}
\text{TE}^n := \frac{1}{2}\sum_{p=1}^N w_p |\bv_p^{n}|^2, \quad 
\text{FE}^n := \left|\frac{\text{TE}^n - \text{TE}^0}{\text{TE}^0}\right|,
\end{equation*}
 where $\text{FE}^n$ represents the relative error with respect to the initial energy. Our schemes conserve energy much better than the forward Euler. In addition, no problematic particles (i.e., particles with imaginary $\Gamma_p^n$) were observed in either the version 1 or version 2 scheme in any of the presented cases. This also explains the similar energy conservation behavior of two schemes shown in Figure~\ref{fig:BM_fracE}. 

\subsection{Spatially inhomogeneous case}

In this subsection, we present results for more physically interesting test cases, including linear and nonlinear Landau damping and the two-stream instability. In each case, different collision frequencies are considered to illustrate the collisional effects. All numerical examples assume one-dimensional physical space and one-dimensional velocity space (1D1V), and the normalized charge is taken as $q=1$.
For performance comparison, we also consider the Boris scheme (or Verlet scheme in the electrostatic case), which is a commonly used time integrator for PIC methods. It captures energy well over the long term, but does not conserve it exactly. In the electrostatic case, the scheme is given as follows:
\begin{equation*}
\begin{aligned}
    \bx_p^* &= \bx_p^n + \frac{\Delta t}{2}\bv_p^n, \\
    \bv_p^{n+1} &= \bv_p^n + \Delta t q \left(\bE (\bx_p^*) + \bv_p^{n+1/2} \times \bB^{\text{ext}}(t^{n+1/2},\bx_p^*)\right) 
                   - \Delta t \nu \bU\left(\bx_p^*, \bv_p^n\right), \\
    \bx_p^{n+1} &= \bx_p^* + \frac{\Delta t}{2} \bv_p^{n+1},
\end{aligned}
\end{equation*}
where $\bv_p^{n+1/2} = (\bv_p^n+\bv_p^{n+1})/2$.  
Unlike our schemes, the field $\bE(\bx_p^*)$ is obtained by solving the Poisson's equation \eqref{Poisson}. In addition, to compare the energy of the Verlet scheme with ours at the same time level, we compute $\bE_h^{n+1}$, although it is not part of the algorithm.

\subsubsection{Linear Landau damping}	

The initial distribution is given by
\begin{align*}
	f_0(x,v) = (1 + 0.1 \cos(kx)) \frac{1}{\sqrt{2\pi}} e^{-v^2/2},
\end{align*}
where $x \in [0, L_x]$ and $k = 0.5$. The size of the $x$-domain is given by $L_x = 2\pi / k$, and the size of the $v$-domain is defined as $L_v = \max_p v_p^0 - \min_p v_p^0$, where $v^0$ denotes the initial velocities. We set the number of grid points in the $x$-domain to $N_x = 100$ and in the $v$-domain to $N_v = 200$. The grid sizes are then determined by $\Delta x = L_x / N_x$ and $\Delta v = L_v / N_v$. The number of particles per physical cell is set to be $P_c = 1.2 \times 10^4$, leading to a total of $N = N_x \times P_c = 1.2 \times 10^6$ particles. The initial particle positions and velocities are generated using random sampling from $f_0(x,v)$. The kernel functions used in both the scheme and the reconstruction are first-degree B-spline $B^1$ with $h=\Delta x$ for the $x$-domain and Gaussian kernel $G$ with $\epsilon = \Delta v$ for the $v$-domain. We choose a time step $\Delta t = 0.01$ and simulate up to the final time $t_{\text{final}} = 15$. 

\begin{figure}[htp!]
    \centering
   \includegraphics[width=\linewidth]{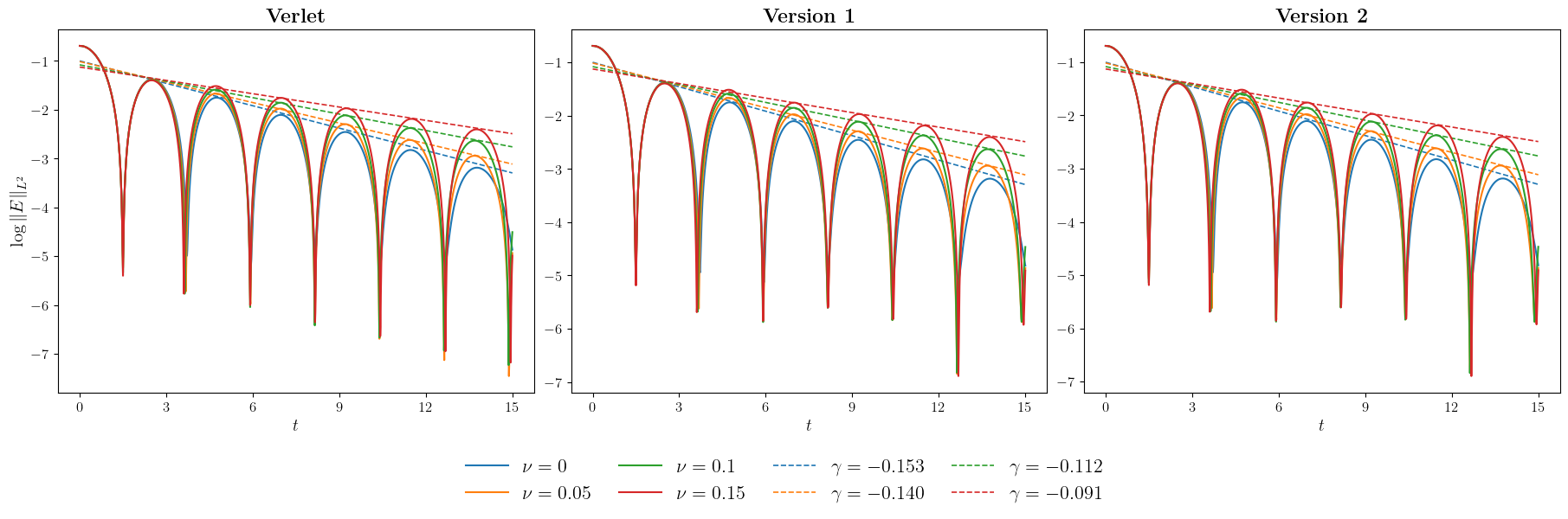}
    \caption{Linear Landau damping test. Electric field energy trace up to time $t_{\text{final}} = 15$.}
    \label{fig:LLD_Etraces}
\end{figure}

Both collisionless and collisional cases are tested, with collision frequencies set to $\nu = 0.05,\ 0.1$, and $0.15$. In the collisionless case, the theoretical damping rate, $-0.1530$, is indicated by the blue dashed line in Figure~\ref{fig:LLD_Etraces}. For collisional cases, the damping rate is computed by performing linear regression on the peaks shown in Figure~\ref{fig:LLD_Etraces}, excluding the first peak. The damping rate exhibits a monotonically decreasing trend as the collision frequency increases. 

\begin{figure}[htp!]
    \centering    \includegraphics[width=\linewidth]{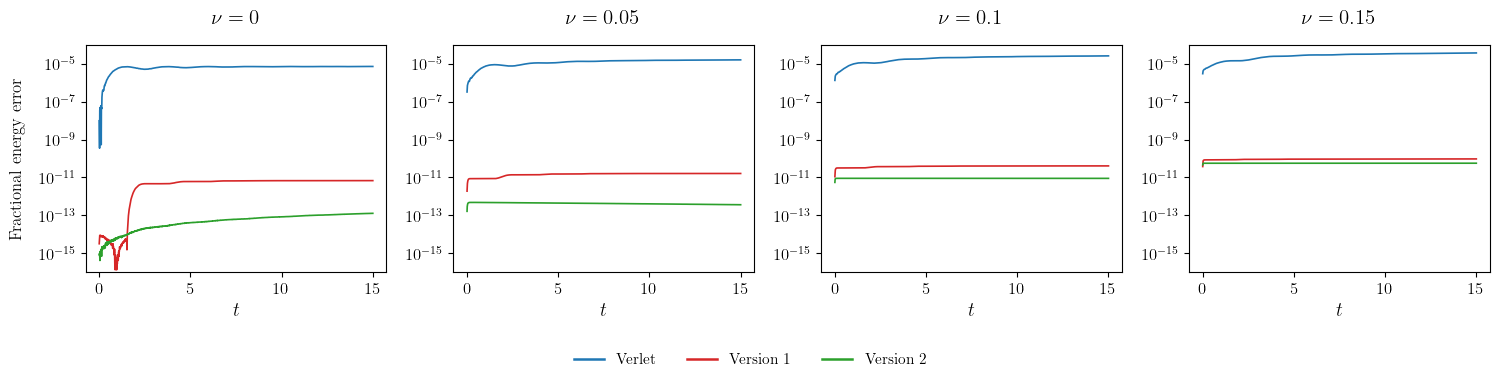}    \includegraphics[width=\linewidth]{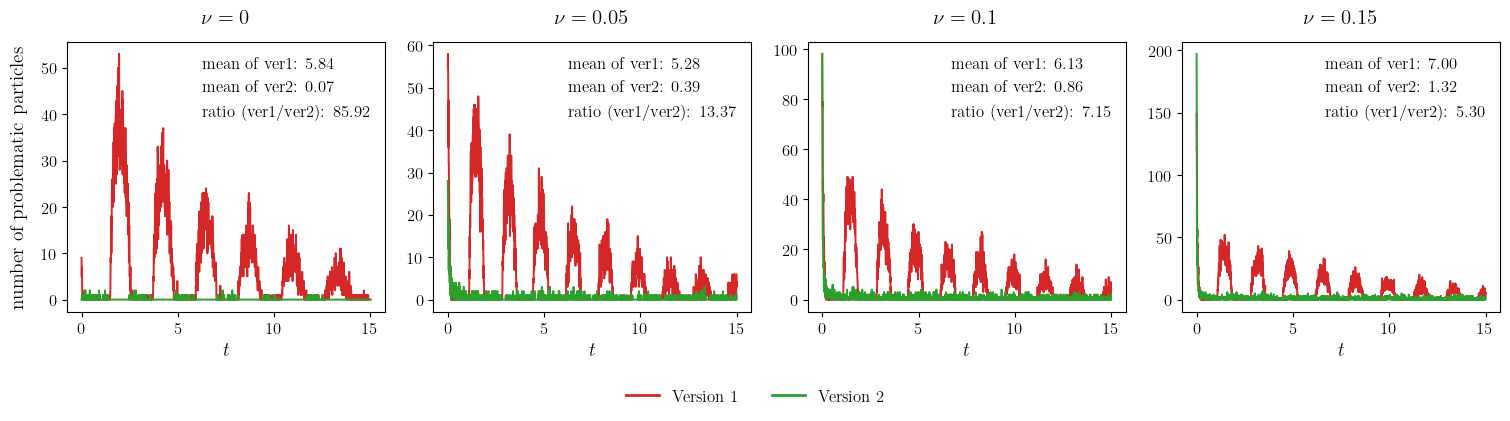}
    \caption{Linear Landau damping test. Top: Fractional change in total energy. Bottom: Number of problematic particles. The means of ver1 and ver2 are computed as the total number of problematic particles divided by the number of time steps. The ratio is defined as the mean of ver1 to the mean of ver2.}
    \label{fig:LLD_fracE_numprobs}
\end{figure}

For energy conservation, we define the total energy and its fractional change as 
\begin{equation*}
\text{TE}^n := \frac{1}{2}\sum_{p=1}^N w_p |\bv_p^{n}|^2 + \frac{h}{2} \sum_h |\bE_h^{n}|^2, \quad 
\text{FE}^n := \left|\frac{\text{TE}^n - \text{TE}^0}{\text{TE}^0}\right|.
\end{equation*}
Figure~\ref{fig:LLD_fracE_numprobs} (top) shows that our schemes preserve the total energy much better than the Verlet scheme, with version 2 outperforming version 1. The difference between version 1 and version 2 arises because problematic particles appear less frequently in version~2, as illustrated in Figure~\ref{fig:LLD_fracE_numprobs} (bottom). It is worth noting that this difference becomes smaller as the collision frequency increases.

\begin{figure}[htp!]
    \centering
\includegraphics[width=\linewidth]{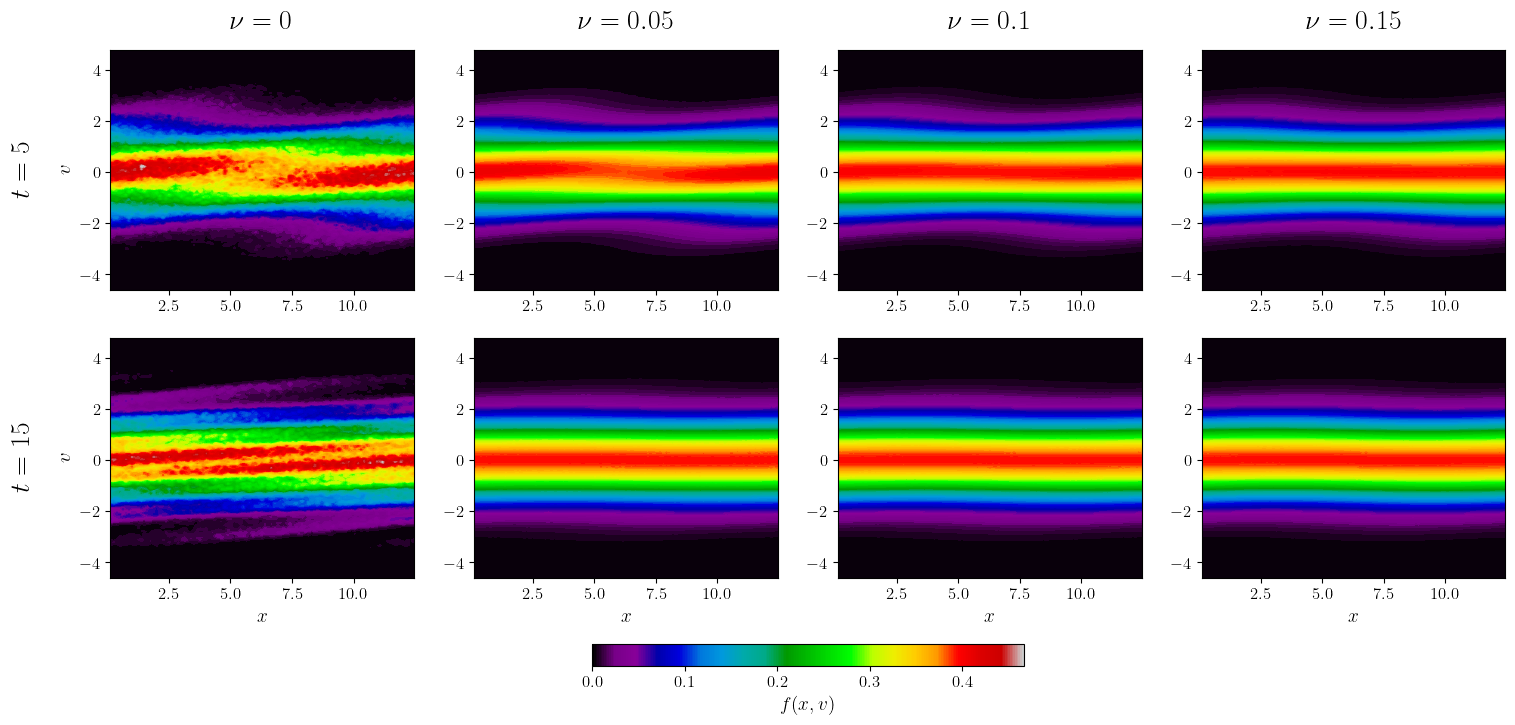}
    \caption{Linear Landau damping test. Time evolution of the phase space distribution at different collision frequencies.}
    \label{fig:LLD_phaseplots}
\end{figure}

Finally, we show the phase space plots in Figure~\ref{fig:LLD_phaseplots} at times $t = 5$ and $t = 15$. Since the results obtained with the Verlet, version 1, and version 2 schemes are visually indistinguishable, we only present the version 1 plots as representative. The distribution function clearly becomes smoother due to collisional effects.

\subsubsection{Nonlinear Landau damping}	

We consider the same initial data as in the linear Landau damping case, but with a stronger perturbation amplitude of $0.5$:
\begin{align*}
	f_0(x,v) = (1 + 0.5 \cos(kx)) \frac{1}{\sqrt{2\pi}} e^{-v^2/2},
\end{align*}
where $x \in [0, 2\pi/k]$ and $k = 0.5.$ The number of grid points, domain sizes, and kernel parameters are the same as in the linear Landau damping case. The number of particles per physical cell is $P_c = 5 \times 10^3$, giving $N = N_x \times P_c = 5 \times 10^5$ total particles. We choose a time step $\Delta t = 0.01$ and simulate up to the final time $t_{\text{final}} = 50$. The collision frequencies are set to $\nu = 0.01,\ 0.05,\ 0.1$.

\begin{figure}[htp!]
    \centering
   \includegraphics[width=\linewidth]{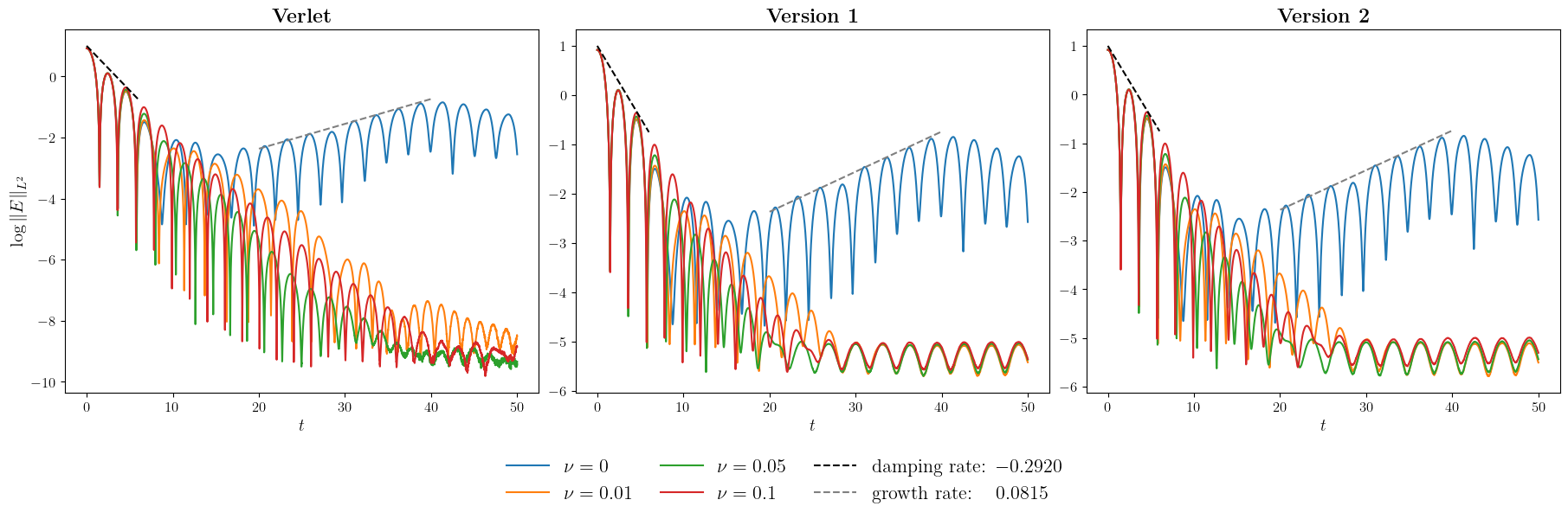}
    \caption{Nonlinear Landau damping test. Electric field energy trace up to time $t_{\text{final}} = 50$.}
    \label{fig:NLD_Etraces}
\end{figure}

Figure~\ref{fig:NLD_Etraces} shows the time evolution of the electric field energy. In the collisionless case, the theoretical damping rate, $-0.2930$, is indicated by the black dashed line and the growth rate,  $0.0815$, by the gray dashed line. With collisions, the electric field energy initially shows a smaller absolute damping rate, similar to the behavior in the linear Landau damping. However, unlike the collisionless case, the field energy does not rebound but instead decays monotonically.

\begin{figure}[htp!]
    \centering
   \includegraphics[width=\linewidth]{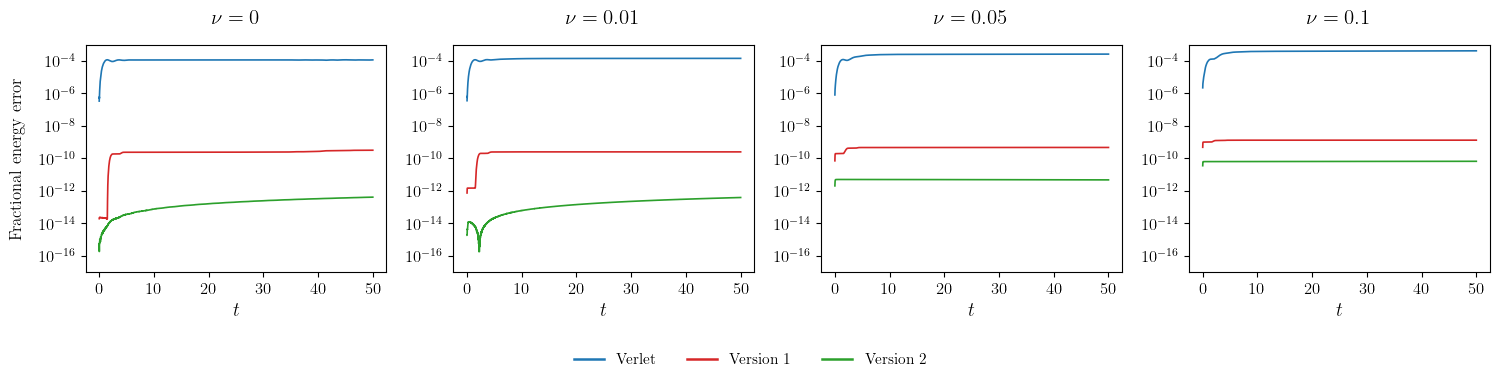}
   \includegraphics[width=\linewidth]{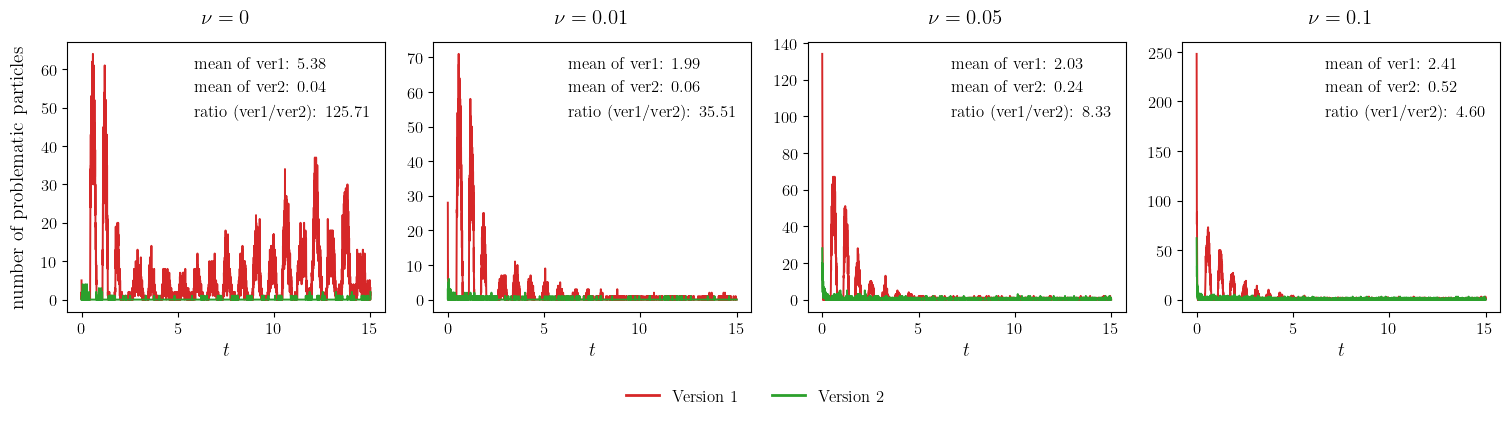}
    \caption{Nonlinear Landau damping test. Top: Fractional change in total energy. Bottom: Number of problematic particles. The mean and ratio are computed in the same way as in Figure 5.
    }
    \label{fig:NLD_fracEs_numprobs}
\end{figure}

In terms of energy conservation, the nonlinear Landau damping case shows similar behaviors to the linear case. Our schemes preserve the total energy much better than the Verlet scheme, with version 2 outperforming version 1, as shown in Figure~\ref{fig:NLD_fracEs_numprobs} (top). Figure~\ref{fig:NLD_fracEs_numprobs} (bottom) further shows that version~2 consistently produces fewer problematic particles than version~1 in all cases.

\begin{figure}[htp!]
    \centering
   \includegraphics[width=\linewidth]{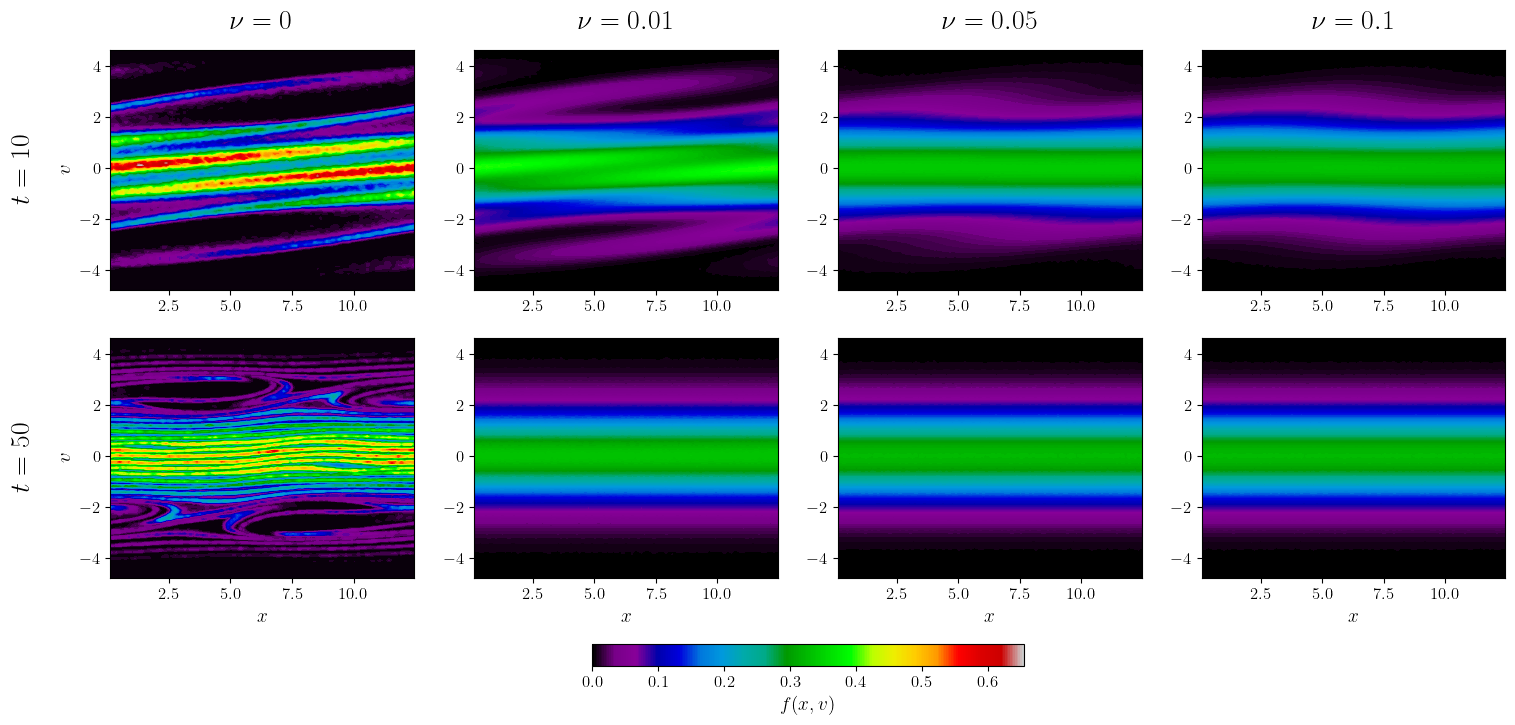}
    \caption{Nonlinear Landau damping test. Time evolution of the phase space distribution at different collision frequencies.
    }
    \label{fig:NLD_phaseplots}
\end{figure}

The collisional effect again drives the distribution toward a Maxwellian, while in the collisionless case the stronger perturbation generates clear filamentations in the phase space over time, as shown in Figure~\ref{fig:NLD_phaseplots}. Since the phase plots from the three schemes are visually indistinguishable, we present only the version 1 plots at $t = 10$ and $t = 50$. 

\subsubsection{Two-stream instability}		

The initial distribution is given by
\begin{align*}
	f_0(x,v) = \left(1 + 0.005 \cos(kx)\right) \frac{1}{2\sqrt{2\pi}} \left( e^{-\frac{(v - 2.4)^2}{2}} + e^{-\frac{(v + 2.4)^2}{2}} \right),
\end{align*}
where $x \in [0, 2\pi/k]$ and $k = 0.2$. The number of grid points, domain sizes, kernel parameters, and particle numbers are the same as in the nonlinear Landau damping case. We set the time step to $\Delta t = 0.1$ and simulate up to the final time $t_{\text{final}} = 50$. The collision frequencies are set to $\nu = 0.001,\ 0.002, \ 0.003$, and $0.004$.

\begin{figure}[htp!]
    \centering    \includegraphics[width=\linewidth]{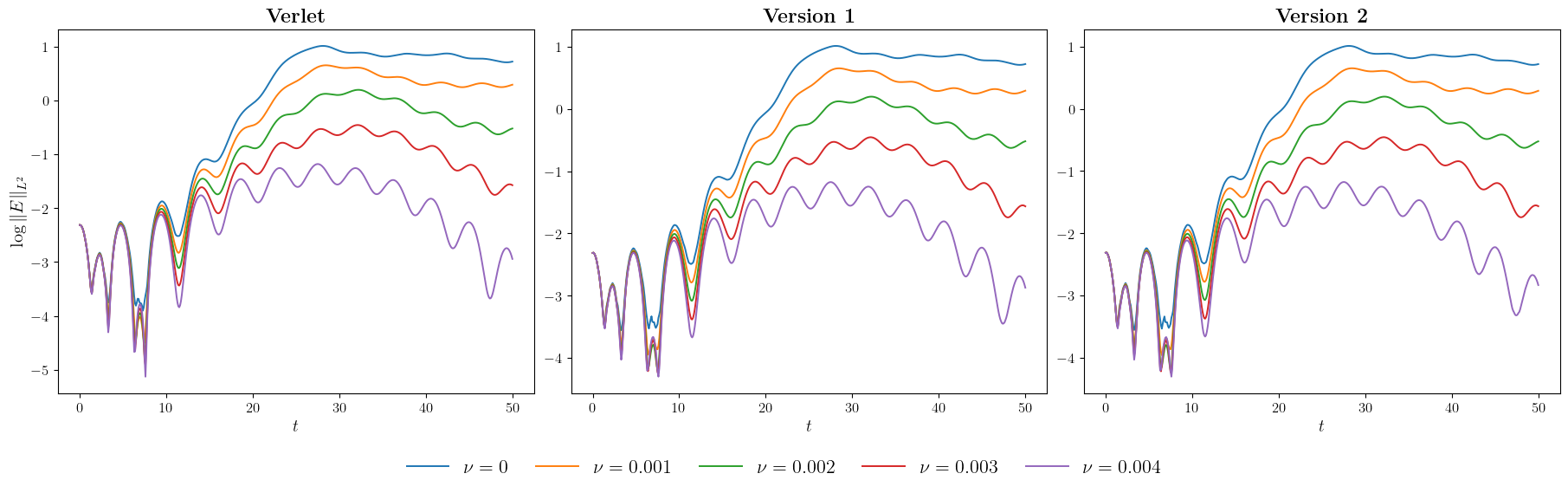}
    \caption{Two-stream instability test. Electric field energy trace up to time $t_{\text{final}} = 50$.}
    \label{fig:TSI_Etraces}
\end{figure}

\begin{figure}[htp!]
    \centering
  \includegraphics[width=\linewidth]{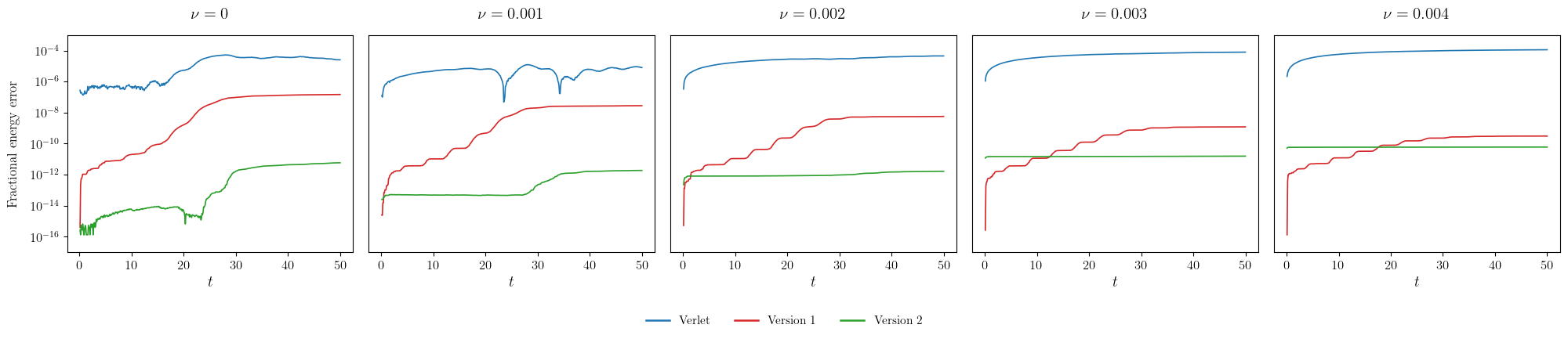}
   \includegraphics[width=\linewidth]{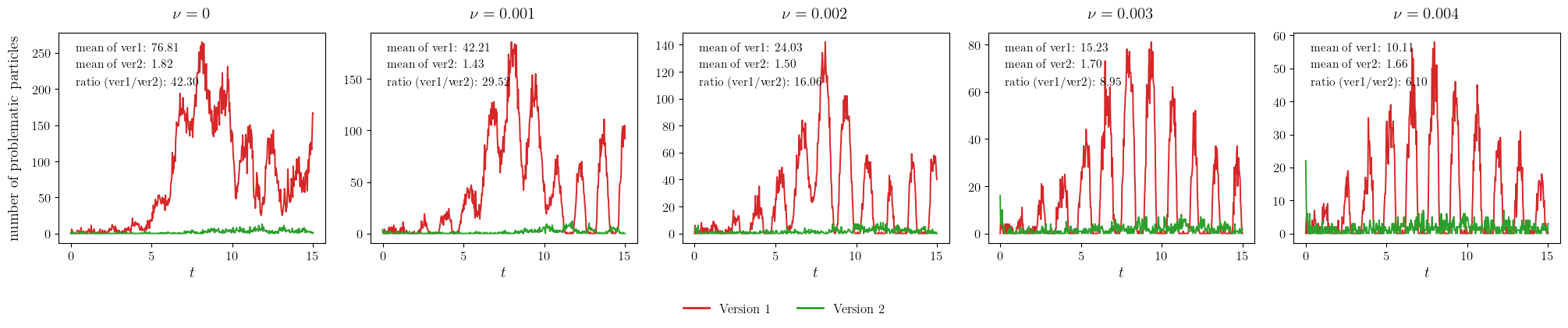}
    \caption{Two-stream instability test. Top: Fractional change in total energy. Bottom: Number of problematic particles. The mean and ratio are computed in the same way as in Figure 5. 
    }
    \label{fig:TSI_fracE_numprobs}
\end{figure}

Figure~\ref{fig:TSI_Etraces} shows the time evolution of the electric field energy. The electric energy decreases as the collision frequency increases.
Figure~\ref{fig:TSI_fracE_numprobs} indicates that versions~1 and~2 provide improved energy conservation compared to the Verlet scheme, and version~2 is more robust than version~1 in producing fewer problematic particles.

\begin{figure}[htp!]
    \centering    \includegraphics[width=\linewidth]{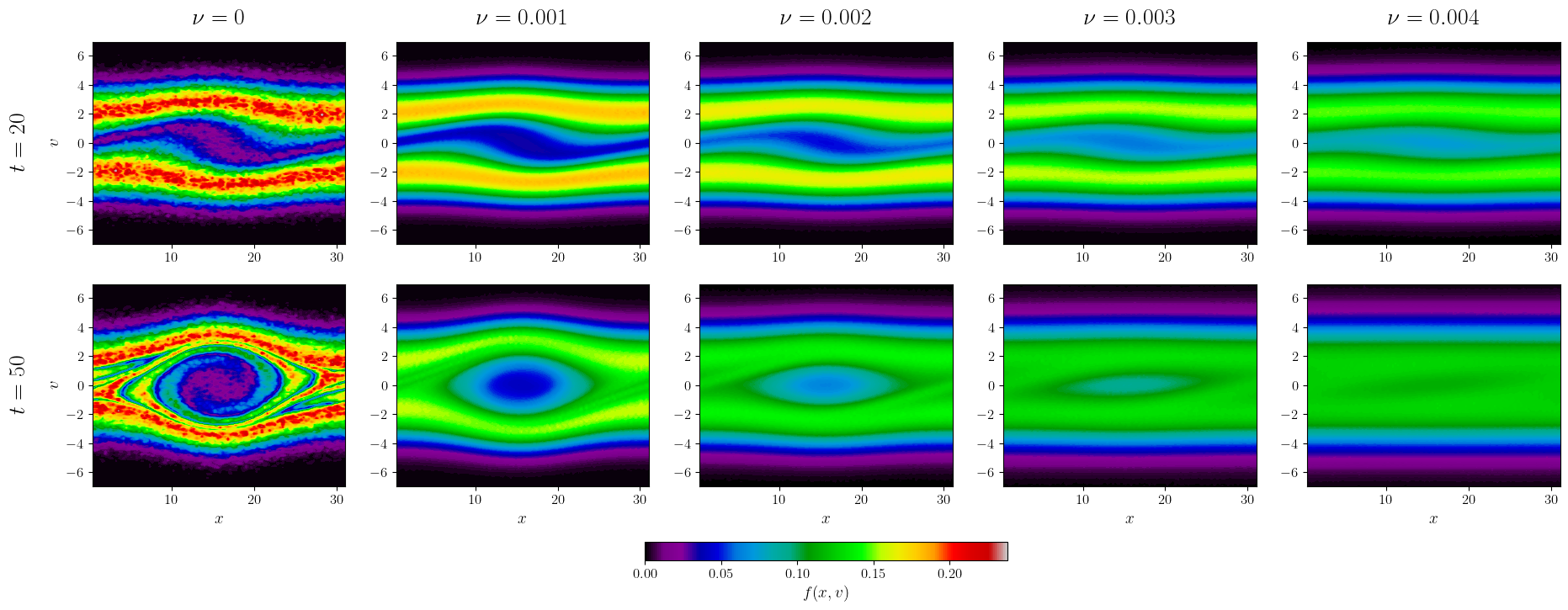}
    \caption{Two-stream instability test. Time evolution of the phase space distribution at different collision frequencies.}
    \label{fig:TSI_phaseplots}
\end{figure}

For the phase plots in Figure~\ref{fig:TSI_phaseplots}, we clearly observe a vortex structure at the center in the collisionless case at $t_{\text{final}} = 50$. As the collision strength increases, this vortex gradually disappears as the distribution approaches the Maxwellian equilibrium more closely. In all collisional cases, the smoothing effect is evident. We present only the version~1 plots at $t = 20$ and $t = 50$, as the three schemes produce visually indistinguishable phase space distributions. 

\section{Conclusion}
\label{Sec:conclusion}

We introduced an explicit, energy-conserving particle method for the Vlasov–Fokker–Planck equation. For the Fokker–Planck collision operator, we proposed a deterministic particle method that can be naturally coupled with the classical (collisionless) PIC method. The conservation property is ensured through a novel optimization procedure. For the resulting particle system, we designed a fully explicit second-order time integrator, 
whose key component is an accuracy-justifiable correction step that guarantees energy conservation. We also presented its extension to the electromagnetic case. The accuracy and energy-conservation properties of the method were demonstrated through a series of benchmark tests. The proposed method can serve as an important tool for studying weakly collisional plasma dynamics, a phenomenon that remains poorly understood both analytically and numerically. Future work includes convergence analysis of the method and further numerical studies in the electromagnetic case.

\section*{Acknowledgement}
This work was partially supported by DOE grant DE-SC0023164. The work of JY and JH was additionally supported by NSF grants DMS-2409858 and IIS-2433957, and DoD MURI grant FA9550-24-1-0254. The work of LR was performed under the auspices of the U.S. Department of Energy by LLNL under Contract DE-AC52-07NA27344.

\bibliographystyle{plain}
\bibliography{hu_bibtex}

	
\end{document}